\documentclass[twocolumn]{aastex631}
\usepackage{amsmath, bold-extra}
\DeclareMathOperator{\asinh}{asinh}
\usepackage{tablefootnote}

\graphicspath{{./}{figures/}}

\begin{document}

\title{Orbital Modeling of \textsc{Hii} 1348B: An Eccentric Young Substellar Companion in the Pleiades}

\correspondingauthor{Gabriel Weible}
\email{gweible@arizona.edu}

\author[0000-0001-8009-8383]{Gabriel Weible}
\affiliation{Department of Astronomy and Steward Observatory, University of Arizona, 933 North Cherry Avenue, Tucson, AZ}

\author[0000-0002-4309-6343]{Kevin Wagner}
\affiliation{Department of Astronomy and Steward Observatory, University of Arizona, 933 North Cherry Avenue, Tucson, AZ}

\author[0000-0003-0454-3718]{Jordan Stone}
\affiliation{U.S. Naval Research Laboratory, 4555 Overlook Ave. S.W. Washington, DC 20375}

\author[0000-0002-2314-7289]{Steve Ertel}
\affiliation{Large Binocular Telescope Observatory, University of Arizona, 933 North Cherry Avenue, Tucson, AZ}
\affiliation{Department of Astronomy and Steward Observatory, University of Arizona, 933 North Cherry Avenue, Tucson, AZ}

\author[0000-0003-3714-5855]{D\'aniel Apai}
\affiliation{Department of Astronomy and Steward Observatory, University of Arizona, 933 North Cherry Avenue, Tucson, AZ}
\affiliation{Lunar and Planetary Laboratory, University of Arizona, 1629 E. University Blvd, Tucson, AZ}

\author[0000-0001-5253-1338]{Kaitlin Kratter}
\affiliation{Department of Astronomy and Steward Observatory, University of Arizona, 933 North Cherry Avenue, Tucson, AZ}

\author[0000-0002-0834-6140]{Jarron Leisenring}
\affiliation{Department of Astronomy and Steward Observatory, University of Arizona, 933 North Cherry Avenue, Tucson, AZ}






\begin{abstract}
Brown dwarfs with known physical properties (e.g., age and mass) are essential for constraining models of the formation and evolution of substellar objects. We present new high-contrast imaging observations of the circumbinary brown dwarf H\textsc{ii} 1348B---one of the few known substellar companions in the Pleiades cluster. We observed the system in the infrared (IR) $L^\prime$ band with the Large Binocular Telescope Interferometer (LBTI) in dual-aperture direct imaging mode (i.e., with the two telescope apertures used separately) on 2019 September 18. The observations attained a high signal-to-noise ratio ($\mathrm{SNR} > 150$) photometric detection and relative astrometry with uncertainties of ${\sim}5 \ \mathrm{mas}$. This work presents the first model of the companion's orbital motion using relative astrometry from five epochs across a total baseline of 23 years. Orbital fits to the compiled data show the companion's semimajor axis to be $a = 140 \substack{+130 \\ -30} \ \mathrm{au}$ with an eccentricity of $e = 0.78 \substack{+0.12 \\ -0.29}$. We infer that H\textsc{ii} 1348B has a mass of $60\text{--}63 \pm 2 \ \mathrm{M_J}$ from comparison to brown dwarf evolutionary models given the well-constrained distance and age of the Pleiades. No other objects were detected in the \textsc{Hii 1348} system, and through synthetic planet injection and retrievals we establish detection limits at a cluster age of $112 \pm 5$ Myr down to ${\sim}10\text{–}30 \ \mathrm{M_J}$ for companions with projected separations of $21.5\text{–}280 \ \text{au}$. With this work, \textsc{Hii} 1348B becomes the second directly imaged substellar companion in the Pleiades with measured orbital motion after HD 23514B.
\end{abstract}




\section{Introduction}
\label{sec:intro}

Nearby wide-orbit substellar companions (roughly speaking, planets and brown dwarfs at $\gtrsim$10 au from stars within a few hundred parsecs) are ideal targets for direct imaging because they are sufficiently removed from the bright glare of their host stars. These objects, and especially those with well-constrained ages, are also important tests for star and planet formation models, as their observed orbits and spectra can be directly observed and compared to predictions  \citep[e.g.,][]{pinfield2006, bowler2020}. Beginning with the work of Johannes Kepler in the early 17th century, elliptical orbits have been fit to astrometry \citep{kepler, kepler_1619}. Despite orbital periods spanning decades to millennia, high-precision astrometry can constrain an imaged companion's orbital properties from just a fraction of an observed orbital arc \citep[e.g.,][]{de-rosa2015, blunt2017, de-rosa2020, bowler2020}. Furthermore, direct photometry and spectroscopy of such substellar objects can reveal population-level trends in atmospheric properties as a function of spectral type, temperature, surface gravity, etc.~ \citep[e.g.,][]{knapp2004, liu2016, sanghi_2023}. Brown dwarf companions, being brighter and more readily directly observable than planets at a given age, can also serve as analogs for giant exoplanets that have similar temperatures and cloud properties---despite plausibly different formation pathways \citep[e.g.,][]{kulkarni1997, burrows2003, faherty_2016}.

As one of the nearest young star clusters, the Pleiades offer an important opportunity for studying how stellar and planetary systems form and evolve. Gas giant planets and brown dwarfs retain heat from the gravitational contraction of formation and then proceed to cool with time \citep[e.g.,][]{baraffe2003, marley2007}. The exact ranges of plausible initial conditions and cooling rates depend on the accretion physics and atmospheric opacities---both of which can be constrained by observations. By measuring the orbits, temperatures, and masses of planets and brown dwarfs at well-known ages, such as in the Pleiades, we can constrain evolutionary models to better understand the formation and evolution of substellar objects.  For example, several studies have illuminated discrepancies between substellar parameters inferred from fitting atmospheric models to spectra and photometry, and those found via comparison to evolutionary models for objects with well-constrained ages \citep[e.g.,][]{dupuy_kraus_2013, zhang_2021, sanghi_2023}. Additionally, as the known sample of directly imaged substellar companions has grown, population-level studies of the eccentricity distribution of substellar objects have hinted at the possibility of distinct formation mechanisms for wide-orbit companions as a function of mass \citep[e.g.,][]{bowler2020, doo2023}.

Open clusters provide an opportunity to study substellar objects at intermediate ages. At $112 \pm 5 \ \mathrm{Myr}$ \citep{dahm2015}, the Pleiades open cluster is older than many young, nearby moving groups and stellar associations. For reference, the age of the $\beta$ Pictoris moving group is $23 \pm 3 \ \mathrm{Myr}$ \citep{mamajek2014}, and the TW Hydrae association is $10 \pm 3 \ \mathrm{Myr}$ old \citep{bell2015}. Among open clusters, the Pleiades is the youngest of the three closest to the Earth. The Hyades is approximately three times closer, but with a significantly older age of ${\sim}800 \ \mathrm{Myr}$ \citep{brandt2015}, and the second-closest,  the Coma Ber open cluster, has an age of $560 \pm 90\ \mathrm{Myr}$ \citep{silaj2014}. Field brown dwarfs span an even wider range of ages up to several Gyr \citep[e.g.,][]{dupuy_liu_2017, best2024}, but often with large uncertainties. Characterizing substellar objects across a wide range of ages is necessary to properly constrain evolutionary models across time, putting the Pleiades in a unique position between very young moving groups/stellar associations and older open clusters/field brown dwarfs.

\begin{figure*}
\plotone{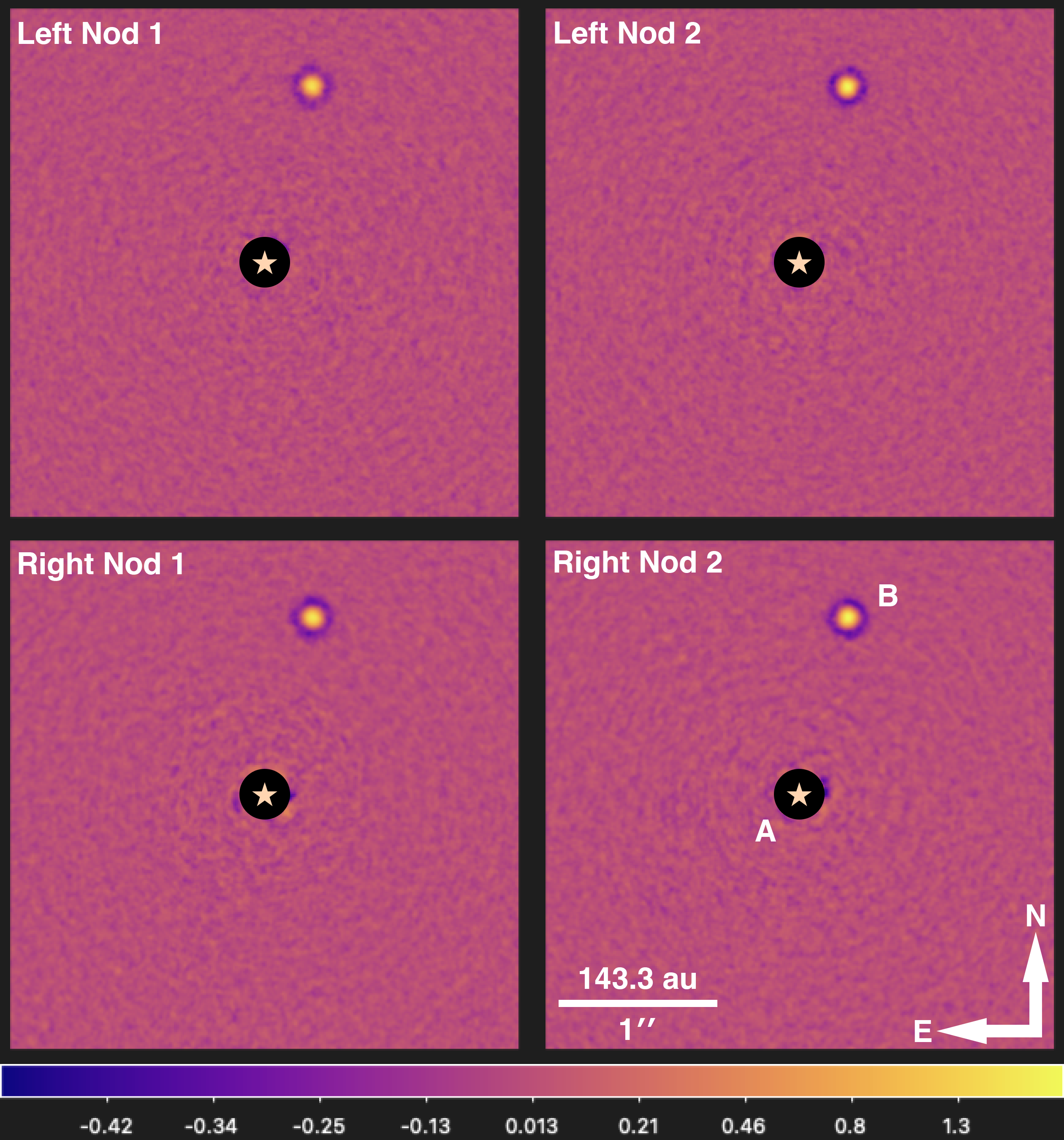}
\caption{ Four ADI--KLIP-subtracted reduced images of the \textsc{Hii} 1348 system, one from each telescope aperture of the LBT at each of two nod positions. The images have been cropped to the inner $3\farcs{1944} \times 3\farcs{1944}$ and are shown with an $\asinh$ stretch in arbitrary units with \texttt{SAOImageDS9} \citep{ds9}. In the lower right panel, the companion is labeled ``B'' and the binary host star residuals are masked and labeled ``A.'' Orientation on the celestial sphere and a scale are shown.
\label{fig:klip}}
\end{figure*}

There are few substellar companions in the Pleiades available for study; thus, each adds important insights into this sparsely known population. At least one brown dwarf companion in the Pleiades has been discovered by radial velocity (RV) observations around the star 2MASS J03432619+2602308 \citep{kounkel2019}, referred to as V* V619 Tau in the Set of Identifications, Measurements, and Bibliography for Astronomical Data \citep[SIMBAD;][]{wenger2000} database. The exoplanet EPIC 210736056 b, detected by transit photometry, orbits the star UCAC4 545-007339 which was determined to have a 62\% probability of membership in the Pleiades \citep{rizzuto2017}. One further exoplanet has been detected by K2-mission transit photometry around a star likely to be a non-member interloper in the Pleiades \citep[K2-77 b,][]{gaidos2017}. More than $200$ isolated brown dwarfs have been identified in the Pleiades \citep{lodieu2012}, beginning with Teide 1, \citep{rebolo1995, rebolo1996}, Calar 3 \citep{rebolo1996}, and Roque 25 \citep{martin1998}. Since the late 1990s, Pleiades brown dwarfs with spectral types ranging from M6 to late L have been discovered and characterized \citep{bejar2018}.

Only three substellar companions to Pleiades member stars have been directly observed: \textsc{Hii} 1348B \citep{metchev2006, geißler2012}, HD 23514B \citep{rodriguez2012}, and \textsc{Hii} 3441B \citep{konishi2016}---all with inferred masses of ${\sim}60 \ \mathrm{M_J}$ from comparison to evolutionary models. These objects provide the best opportunities for joint orbital and atmospheric characterization of substellar companions in the Pleiades cluster. In this work, we provide new insights into the circumbinary brown dwarf \textsc{Hii} 1348B to better constrain its orbital and atmospheric properties---thus improving what is known of the small population of substellar companions in the Pleiades.

\textsc{Hii} 1348B is a wide-orbit, ${\sim}60 \ \mathrm{M_J}$ companion to a K-type double-lined spectroscopic binary (SB2) star \citep{geißler2012}. The discovery of \textsc{Hii} 1348B as a substellar companion was reported by \cite{metchev2006} and \cite{geißler2012} after previously being regarded as a background star unrelated to the \textsc{Hii} 1348 system by \cite{bouvier1997}. Follow-up observations of the system as part of the Strategic Explorations of Exoplanets and Disks with Subaru \citep[SEEDS;][]{SEEDS2009} were presented by \cite{yamamoto2013}. As an imaged companion, there is the opportunity for astrometric orbital modeling of \textsc{Hii} 1348B in addition to spectroscopic and photometric studies. Importantly, the inferred orbital elements for this companion  might be linked to its formation history and past dynamical interactions \citep[e.g.,][]{veras2009, bowler2020}.

Such directly imaged wide-orbit companions more massive than ${\sim}10\text{–}20 \ \mathrm{M_J}$ are not likely to be formed in-situ by core accretion \citep[CA;][]{mordasini2009I, mordasini2009II, wagner2019}. Consistent with this picture, and given that \textsc{Hii} 1348B is on a circumbinary orbit around a binary star, fragmentation of a single molecular cloud creating both stellar and substellar bodies is one compelling explanation for the system's present configuration. Alternatively, if the brown dwarf formed after the stellar components via disk fragmentation \citep[e.g.,][]{cameron1978, boss1997, kratter2010}, then dynamical scattering with the inner binary could also account for the wide orbit of \textsc{Hii} 1348B \citep[e.g.,][]{heggie1975, reipurth2001, veras2009}. In either case---formation via cloud or disk fragmentation---a relatively large eccentricity may be expected for such massive brown-dwarf companions \citep[e.g.,][]{bowler2020, nagpal2023, doo2023}.

Substellar companions formed in protoplanetary disks via CA are likely to have an eccentricity distribution that is lower than that of companions formed via fragmentation \citep[e.g.,][]{ribas_2007}. Uncovering the orbital eccentricity of \textsc{Hii} 1348B is thus a major goal of our present study. Simultaneously, we aim to establish an independent mass estimate for \textsc{Hii} 1348B by comparing the companion's first-reported $L^\prime$-band brightness to evolutionary models---also to be linked to the system's formation history. Mass estimates from an inferred bolometric luminosity for the companion are also computed. Additionally, we fit atmospheric models to the available broadband photometry of \textsc{Hii} 1348B to infer its bulk atmospheric properties. Finally, these observations place constraints on any additional, less massive companions within the system.

\section{Observations and Data Reduction} \label{sec:obs_data}
\subsection{Observations}
We observed the \textsc{Hii} 1348 system on 2019 September 18  with the $L$- and $M$-band InfraRed Camera \citep[LMIRCam;][]{leisenring2012} inside the Large Binocular Telescope Interferometer \citep[LBTI;][]{hinz2016, ertel2020}. The observations were non-interferometric, with the two 8.4 m telescope apertures being imaged separately on different areas of the detector.  The system was imaged with the standard $L^\prime$ broadband filter on LMIRCam (``Std-L'', $\lambda_\mathrm{c} = 3.70 \ \mu\mathrm{m}$, $\Delta \lambda_\text{FWHM} = 0.58 \ \mu\mathrm{m}$).\footnote{The filter's transmission curve, measured at 77 K, is provided at: \url{http://svo2.cab.inta-csic.es/theory/fps/index.php?id=LBT/LMIRCam.L\_77K\&\&mode=browse\&gname=LBT\&gname2=LMIRCam\#filter}} The LBTI's adaptive secondary mirrors (one for each aperture) and high-performance adaptive optics (AO) systems \citep{esposito2010, bailey2014, pinna2016} were used to correct wavefront errors from atmospheric distortion. Nodding was performed by periodically moving the telescope between two positions to allow for the subtraction of thermal background radiation from the sky and telescope. Either 200 or 400 images were captured in each nod position before the telescope was moved, with the change being made mid-observation to minimize overhead. A total of 11,016 images were taken with an integration time of 0.9065 seconds each, resulting in a total integration of $2.77 \ \mathrm{hr}$.

The spatially unresolved binary host star is unsaturated in our frames, which enables better astrometric and photometric accuracy as a trade-off for higher read noise.  However, the mid-IR photon noise from the thermal background dominates over the read noise of the $2048 \times 2048$ HAWAII-2RG (H2RG) detector installed in LMIRCam \citep{leisenring2012}. $124^\circ$ of total field rotation were covered during the observing sequence, allowing for differential imaging techniques to be applied for star subtraction (see Section \ref{ssec:reduction}).

\subsection{Data Reduction}
\label{ssec:reduction}
Our images were co-added by taking the means of each set of 25 consecutive frames, after performing correlated double sampling (CDS) on each image to remove $kTC$ reset noise. The 200 or 400 images taken at each nod position resulted in 8 or 16 coadded frames per nod. Bad pixels with values outside of $\pm 5$ standard deviations from the mean of their surrounding pixels were identified and masked. We then reduced the thermal background radiation in each image by subtracting the mean of the frames from a neighboring nod position. Optical distortion was corrected using the distortion solutions provided by routine instrument calibrations for each side of the telescope.\footnote{\url{https://scienceops.lbto.org/lbti/data-retrieval-reduction-publication/distortion-correction-and-astrometric-solution/}} (Left/SX and Right/DX) separately \citep{maire2015, spalding2019}

We split our observations into four semi-independent data sets: one for each side of the LBT at each nod position.  The two sides of the telescope each have only three warm optics: their primary, (adaptive) secondary, and tertiary mirrors. Light from the two tertiaries is then directed into either side of the liquid nitrogen--cooled LBTI beam combiner with unique optical components for either aperture upstream of the Nulling and Imaging Camera (NIC) cryostat \citep{hinz2016}. Inside of NIC, the two beams share the same optics downstream of the ``roof mirrors''---including a trichroic which transmits the $3\text{--}5 \ \mu\mathrm{m}$ light to LMIRCam. LMIRCam re-images a $10'' \times 10''$ field and transmits the beams through the $L^\prime$ filter before focusing onto the H2RG detector \citep{leisenring2012}. Through the shared optics, the two beams follow different optical paths, providing semi-independence. Similarly, the two nod positions do not provide complete independence, instead directly varying the optical paths taken by the light of both beams through the telescope and science instruments. Separating our data into four subsets in this manner has allowed us to obtain four astrometric and photometric measurements that we have checked for agreement, and subsequently combined through a mean.

The images across the four data sets were cropped to $500\ \text{px} \times 500\ \text{px}$ (${\sim}5\farcs{3} \times 5\farcs{3}$) regions aligned to a common center by cross-correlation with a reference frame. In this process, the location of the maximum correlation between the reference frame and the frame being centered gives the relative shift of the two stellar point spread functions (PSFs), such that they can be aligned. Sub-pixel alignment was achieved using bi-linear interpolation. Each centered frame was then cross-correlated with the median frame of our observations, with images having maximum cross-correlations of less than 0.994 being rejected. We iterated upon this frame-selection threshold to maximize the signal-to-noise ratio (SNR) on \textsc{Hii} 1348B, with 19.8\% of frames being rejected. Images that were rejected by the maximum cross-correlation threshold include those that were taken when the AO control loop opened or the correction was otherwise degraded, making them not useful for high-contrast imaging and precision astrometry. We also accounted for detector bias in each image (vertical and horizontal stripes) by subtracting from each row and column the mode of its counts.

The binary host star's (spatially unresolved) PSF was subtracted using either a standard implementation of classical angular differential imaging \citep[cADI;][]{marois2006} or ADI together with Karhunen-Loève Image Projection \citep[ADI--KLIP:][]{soummer2012}. The LBTI does not have a field derotator, so taking the median over all images statistically rejects any companions or background sources. Over the full observing sequence, sources outside of the centered target star appear to rotate through arcs across many pixels, covering each for only a small fraction of the total integration. Therefore, their smeared signals are statistically rejected by the median---leaving only the central stellar PSF. This is the core requirement for cADI. The KLIP-ADI algorithm performs essentially the same function as cADI, but it allows for the construction of unique stellar PSF models for each frame in the observing sequence via projection onto a (truncated) set of eigenimages derived from surrounding frames taken at sufficiently distant parallactic angles \citep{soummer2012}.

The cADI-subtracted images yield slightly lower SNR on the companion than with KLIP-ADI but allow for faster astrometric/photometric grid searches that are described in Section \ref{ssec:astrometry}. KLIP parameters were optimized through trial and error to maximize the SNR recovered for \textsc{Hii} 1348B. Our KLIP algorithm was applied independently to each half of four annuli for a total of eight segments per image. As part of our KLIP algorithm, frames were binned by taking the mean of every three coadded frames, for an overall KLIP co-adding of 75 (before ADI--KLIP star subtraction we had already coadded the frames by 25). Seven KLIP components were retained to create the PSF models used for stellar subtraction. For each of the four data sets, we derotated the images to north-up and used a noise-weighted mean to combine the ADI--KLIP-subtracted frames \citep{bottom2017}. Our four ADI--KLIP images are shown in Figure \ref{fig:klip}. The dark rings around the companion are artifacts of a high-pass filter, where we subtracted from our reduced images smoothed versions of themselves to reduce low--spatial frequency noise, applied before the PSF subtraction and image combination with cADI or ADI--KLIP. This high-pass filter does not affect our reported astrometric or photometric measurements due to the forward modeling of negative sources injected into the data as detailed in Section \ref{ssec:astrometry}.

\begin{figure*}
\plotone{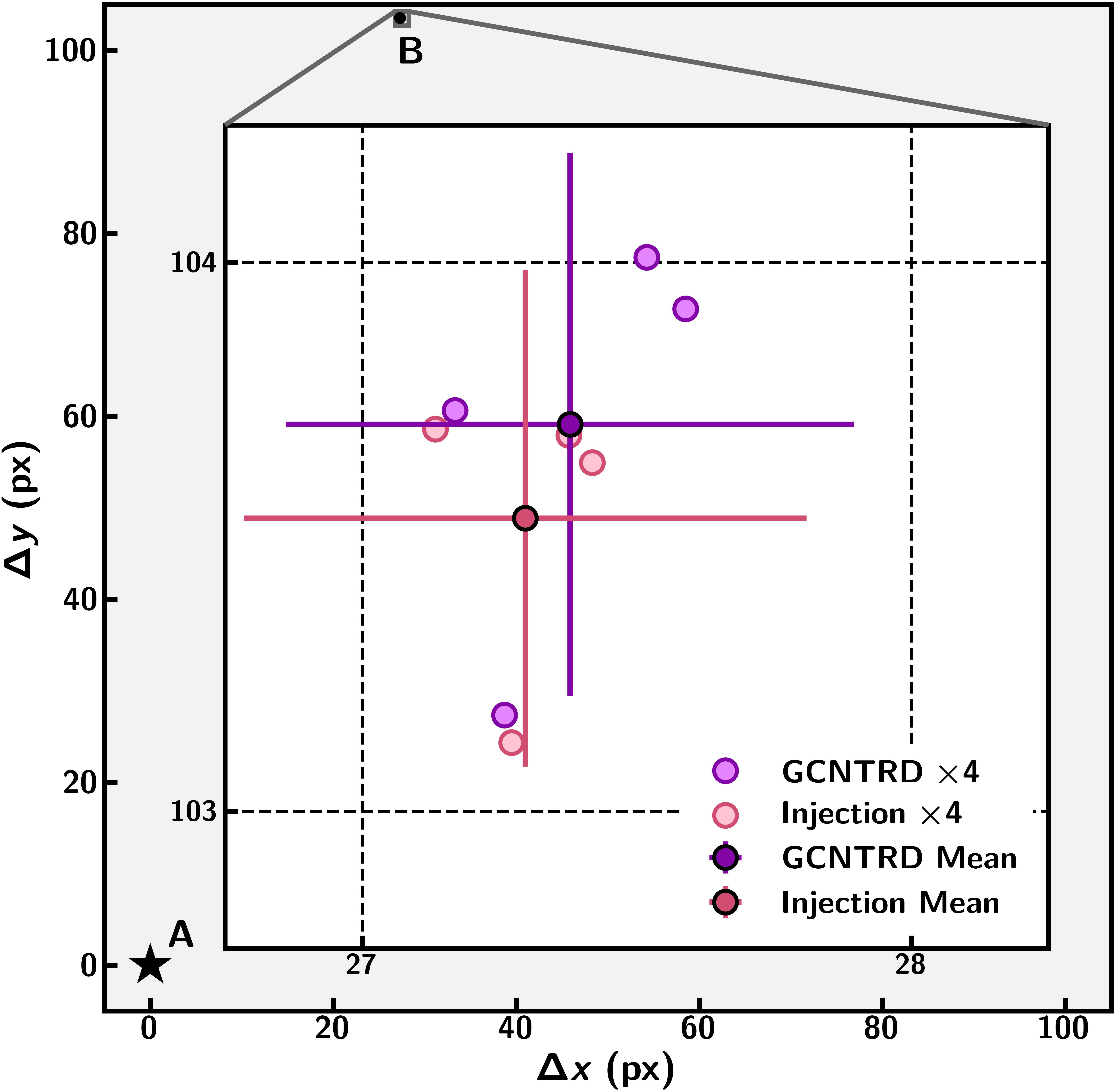}
\caption{Detector-plane (pixels) plot of relative astrometry performed for the \textsc{Hii} 1348 system (see Section \ref{ssec:astrometry}). The dashed lines indicate LMIRCam pixel boundaries after aligning images to a common center. The fuchsia points are the measurements obtained from negative artificial companion injections into each of the four data sets, with their mean value shown with a black outline and error bars. The purple points are the measurements obtained for each data set using the IDL \texttt{GCNTRD} program, with their mean value shown with a black outline and error bars. Inferred error bars on the four individual measurements with either technique are not shown for readability.  The mean astrometric measurements and their uncertainties were computed as the inverse-variance weighted mean, where the variance was inflated by a factor of 2 to account for the only semi-independence between data sets (see Section \ref{sssec:astrometric_uncert}).
\label{fig:quad_astro}}
\end{figure*}

\subsubsection{Astrometric Solutions and Distortion Correction}
\label{sssec:solns}
We used two independent true-north and plate scale solutions, one for each aperture of the LBT, to rotate our frames north-up and match their plate scales. Astrometric solutions were yielded by standard observations of the Orion Trapezium, corrected for distortion with the same SX and DX distortion solutions applied to our observations of \textsc{Hii} 1348. Measured stellar baselines were compared to LBT catalog astrometry of the Trapezium published by \cite{close2012}. The plate scales determined were $10.648\substack{+0.039 \\ -0.050}  \ \mathrm{mas} \ \mathrm{px}^{-1}$ and $10.700\substack{+0.042 \\ -0.051} \ \mathrm{mas} \ \mathrm{px}^{-1}$ for the SX and DX apertures, respectively. The higher uncertainty in each case was taken as a symmetric Gaussian error when propagated through our error analysis.

The catalog positions were not corrected for differential stellar proper motions---the absolute proper motions in the Trapezium are expected to be around $1.5  \ \mathrm{mas} \ \mathrm{yr}^{-1}$ \citep{close2012}. The proper motions of stars in the Orion Nebula Cluster are largely random, consistent with a normal distribution having only a slight elongation along the Orion filament \citep{kim2019}. Therefore, there will not be a significant systematic error introduced by neglecting the stars' proper motions since the catalog astrometry was reported. However, the random errors on the true-north and plate scale solutions are higher than if similar catalog astrometry of the Orion Trapezium observed at a later epoch were referenced.

The observed difference in plate scale between the two apertures is measured after distortion corrections are applied. So, it is likely largely a result of the specific distortion solutions chosen instead of optical differences between the two apertures. The astrometric solutions also provided true-north rotation angles for each aperture, requiring the SX images to be rotated $-1.278^\circ\substack{+0.131^\circ \\ -0.225^\circ}$ East of North, and the DX images to be rotated $1.001^\circ\substack{+0.254^\circ \\ -0.237^\circ}$ East of North. The higher uncertainty for each value was taken as a symmetric Gaussian error when propagated. The difference in orientation between the SX and DX images of ${\sim}2^{\circ}$  has been consistently measured across all recent astrometric solutions and is thus found to be a real difference in orientation, i.e., a pupil rotation. Rotation to true-north was performed using bilinear interpolation with the IDL \texttt{ROT} function. The DX images were magnified to the same nominal plate scale as the SX images ($10.648 \ \text{mas} \ \text{px}^{-1}$) after stellar PSF subtraction, using the IDL \texttt{CONGRID} function with linear interpolation.


\begin{deluxetable*}{lllr}
\tablewidth{0pt}
\tablecaption{Relative Astrometry of \textsc{Hii} 1348B\label{tab:astrometry}}
\tablehead{\colhead{Epoch} & \colhead{Separation Angle ($''$)} & \colhead{PA (\textdegree)} & \colhead{\hspace{.75cm}Reference}\hspace{0.75cm}}
\startdata
1996-09-25 to 1996-10-01   &   $1.09 \pm 0.02$     &   $347.9 \pm 0.7$ &   \cite{bouvier1997}\\
2004-10-03          &   $1.097 \pm 0.005$   &   $346.8 \pm 0.2$ &   \cite{geißler2012}\\
2005-11-21          &   $1.12 \pm 0.02$     &   $346.8 \pm 0.6$ &   \cite{geißler2012}\\
2011-12-23          &   $1.12 \pm 0.03$     &   $346.1 \pm 0.9$ &   \cite{yamamoto2013}\\
2019-09-18          &   $1.140 \pm 0.005$   &   $345.2 \pm 0.3$ &   This Work\\
\enddata
\tablecomments{\cite{bouvier1997} reported a range of observation dates and did not list astrometric uncertainties. We assumed the approximate midpoint of the range of dates as 50354 MJD and the astrometric uncertainties as estimated by \cite{geißler2012}.}
\end{deluxetable*}

\section{Results}
\label{sec:results}
\subsection{Relative Astrometry}
\label{ssec:astrometry}
We performed high-resolution relative astrometry with LBTI to inform the first model of the orbit of \textsc{Hii} 1348B. We used two methods to measure the relative position of this companion as a check for consistency. The first method involved injecting negative sources into our data to best negate the signal from \textsc{Hii} 1348B \citep[e.g.,][]{lagrange2010, marois2010, bonnefoy2011, apai2016}. Negative source injections were performed to iteratively minimize the residuals (standard deviation of pixel values) in the region of the observed position of \textsc{Hii} 1348B in a 3-D grid search over $(x, y)$ pixel coordinates and the (negative) contrast of the injected source. The PSF model for the injected negative source was taken from the median frame of our observations, mentioned previously in Section \ref{ssec:reduction} for use with an image-rejecting cross-correlation threshold. The injections were performed after frame selection but before stellar PSF subtraction. This allowed for the negative injections to be forward modeled through the differential-imaging cADI star subtraction algorithm (see Section \ref{ssec:reduction}).
 
For the astrometric and photometric measurements of \textsc{Hii} 1348B, cADI was used over ADI--KLIP to subtract the binary star's unresolved PSF to reduce the computation time required for the grid search. This enabled a finer astrometric and photometric grid search in less time than reasonably attainable with ADI--KLIP, with a limited reduction in SNR. The SNR of \textsc{Hii} 1348B attained with our cADI processing is ${\sim}8\%$ lower than that obtained with ADI--KLIP. This marginally lower SNR is not expected to significantly impact the astrometric measurements, with $\mathrm{SNR} \gtrsim 70$ achieved in each of the four data sets. We performed iteratively finer grid searches for each of the four data sets to yield four semi-independent $L^\prime$-band astrometric and photometric measurements of \textsc{Hii} 1348B.

The second astrometric measurement technique involved fitting 1-D Gaussian distributions to the marginalized $x$- and $y$-profiles of the PSF of \textsc{Hii} 1348B using the \texttt{GCNTRD} procedure in IDL. \texttt{GCNTRD} uses the DAOPHOT \citep{daophot} \texttt{FIND} centroid algorithm and it is found in the IDL Astronomy User's Library\footnote{\url{https://github.com/wlandsman/IDLAstro}}. The four $(x,y)$ coordinates of the best-fitting pairs of 1-D Gaussian profiles to the companion's PSF in each of the four data sets' reduced images then served as an additional set of semi-independent astrometric (but not also photometric) measurements.

\begin{deluxetable*}{llr}
\tablewidth{0pt}
\tablecaption{\textsc{Hii} 1348 System Parameters\label{tab:system}}
\tablehead{\colhead{Parameter} & \colhead{Value} & \colhead{Reference}}
\startdata
R.A.~(h m s) &   	03 47 18.060599 &   \cite{gaiaedr3}\\
Decl.~(d m s) &   +24 23 26.76373 &   \cite{gaiaedr3}\\
$\pi \ (\mathrm{mas})$   &   $6.979 \pm 0.0305$ &   \cite{gaiaedr3}\\
$d \ (\mathrm{pc})$             &   $143.3 \pm 0.6$   &   \cite{gaiaedr3}\\
$\mathrm{Age} \ (\mathrm{Myr})$  &   $112 \pm 5$    &   \cite{dahm2015}\\
\hline
\multicolumn{3}{c}{\textsc{Hii 1348A} (a \& b)}\\
\hline
Spectral Type   &   K5 V    &   \cite{bouvier1997}\\
$B$ &   $13.92 \pm 0.05$    &   \cite{zacharias2012}\\
$V$ &   $12.81 \pm 0.05$    &   \cite{zacharias2012}\\
$W1$   &   $9.615 \pm 0.023$               &   \cite{cutri2014}\\
$(B-V)_\mathrm{a}$ &   $1.05$  &   \cite{queloz1998}\\
$(B-V)_\mathrm{b}$ &   $1.35$  &   \cite{queloz1998}\\
$P_\mathrm{orb} \ (\mathrm{d})$    &   $94.805 \pm 0.012$   &   \cite{torres2021}\\
$\gamma \ (\mathrm{km} \ \mathrm{s}^{-1})$    &   $6.367 \pm 0.022$   &   \cite{torres2021}\\
$K_\mathrm{a} \ (\mathrm{km} \ \mathrm{s}^{-1})$  &   $20.163 \pm 0.054$  &   \cite{torres2021}\\
$K_\mathrm{b} \ (\mathrm{km} \ \mathrm{s}^{-1})$  &   $25.85 \pm 0.334$  &   \cite{torres2021}\\
$a_\mathrm{ab}\sin{i} \ (\mathrm{R_\odot})$   &   $71.77 \pm 0.51$   &   \cite{torres2021}\\
$a_\mathrm{a}\sin{i} \ (10^6 \ \text{km})$  &   $21.878 \pm 0.046$  &   \cite{torres2021}\\
$a_\mathrm{b}\sin{i} \ (10^6 \ \text{km})$    &   $28.05 \pm 0.35$    &   \cite{torres2021}\\
$e$ &   $0.5543 \pm 0.0017$ &   \cite{torres2021}\\
$\omega_\mathrm{a} \ \text{(\textdegree)}$  &   $82.20 \pm 0.32$    &   \cite{torres2021}\\
$T_0 \ (\text{HJD, 2,400,000+})$    &   $\text{56,452.796} \pm 0.068$   &   \cite{torres2021}\\
$M_\mathrm{ab} \ (\mathrm{M_\odot})$    &   $1.22 \pm 0.09$ &   \cite{geißler2012}\\
$M_\mathrm{a} \ (\mathrm{M_\odot})$ &   $0.67 \pm 0.07$  &   \cite{geißler2012}\\
$M_\mathrm{b} \ (\mathrm{M_\odot})$ &   $0.55 \pm 0.05$  &   \cite{geißler2012}\\
$M_\mathrm{a} \sin^3{i} \ (\mathrm{M_\odot})$   &   $0.3101 \pm 0.0083$ &   \cite{torres2021}\\
$M_\mathrm{b} \sin^3{i} \ (\mathrm{M_\odot})$   &   $0.2418 \pm 0.0036$ &   \cite{torres2021}\\
$q$ &   $0.7799 \pm 0.0098$\tablenotemark{a} &   \cite{torres2021}\\
\hline
\multicolumn{3}{c}{\textsc{Hii 1348B}}\\
\hline
Spectral Type   &   $\mathrm{M8} \pm 1$  &   \cite{geißler2012}\\
$J$     &   $16.04 \pm 0.09$    &   \cite{geißler2012}\\
$H$     &   $15.30 \pm 0.09$    &   \cite{geißler2012}\\
$H$     &   $15.7 \pm 0.4$    &   \cite{yamamoto2013}\\
$K_S$   &   $14.88 \pm 0.09$    &   \cite{geißler2012}\\
$K$   &   $15.06 \pm 0.05$    &   \cite{bouvier1997}\\
$L^\prime$     &   $14.65 \pm 0.06$      &   This Work\\
$a \ (\text{au})$   &   $140 \substack{+130 \\ -30}$  &   This Work\\
$e$ &   $0.78 \substack{+0.12 \\ -0.29}$    &   This Work\\
$i \ (\text{\textdegree})$  &   $115 \substack{+25 \\ -9}$  &   This Work\\
$\omega \ (\text{\textdegree})$ &   $(13 \ \mathrm{or} \ 193) \substack{+70 \\ -36}$  &   This Work\\
$\Omega \ (\text{\textdegree})$ &   $(141 \ \mathrm{or} \ 321) \substack{+26 \\ -22}$ &   This Work\\
$\tau$  &   $0.85 \substack{+0.11 \\ -0.07}$ &   This Work\\
$M \ (\mathrm{M_J})$   &   $60\text{--}63 \pm 2$  &   This Work\\
\enddata
\tablenotetext{a}{Mass ratio of \textsc{Hii} 1348Ab to \textsc{Hii} 1348Aa.}
\end{deluxetable*}

 \subsubsection{Astrometric Error Propagation}
 \label{sssec:astrometric_uncert}
 For both the IDL \texttt{GCNTRD} and negative-injection astrometric measurement techniques, mean positions with standard errors on the mean were computed. The resulting mean relative astrometry from negative injections in on-sky polar coordinates is reported in Table \ref{tab:astrometry}, and the measurements made with both negative injections and the IDL \texttt{GCNTRD} procedure are plotted in the common-center, SX-aperture detector plane in Figure \ref{fig:quad_astro}. The sources of uncertainty considered in this work on each of four measurements with a given measurement technique (``Injection $\times$ 4'' and ``GCNTRD $\times$ 4'' in Figure \ref{fig:quad_astro}---inferred error bars are not shown for legibility) are:
\begin{enumerate}
    \item The sample standard deviation of the four data sets' measured positions with a given measurement technique (in projected separation, $\rho$, and position angle, PA)
    \item The reference catalog astrometric uncertainties (in $\rho$ and PA)
    \item The uncertainty in plate scale for the given aperture of LBTI/LMIRCam used to convert between pixel coordinates and on-sky coordinates ($\rho$ only)
    \item The uncertainty in true-north rotation angle for the given aperture of LBTI/LMIRCam (PA only)
\end{enumerate}
All sources of uncertainty on a single astrometric measurement were assumed to be Gaussian and independent.

 We assume that the sample standard deviation of four measurements of the position of \textsc{Hii} 1348B for either astrometric technique includes within itself:
 \begin{enumerate}
     \item The measurement uncertainty for the astrometric technique used (negative injections or \texttt{GCNTRD})
     \item The uncertainty in centering the spatially unresolved host binary's PSF
     \item The astrometric uncertainty remaining from residual distortions in the reduced images after distortion correction
 \end{enumerate}
 We, therefore, do not make any assumptions on the magnitudes of these three sub-sources of error individually and only consider their combined effect on the spread of measurements made across telescope apertures and nodding positions. The sample standard deviations of the four nominal projected separation measurements across both apertures and nod positions for the \text{GCNTRD} and negative injection techniques are $0.40$ pixels and $0.27$ pixels on the LMIRCam detector, respectively. The sample standard deviations of the PA measurements are $0.081^\circ$ for \texttt{GCNTRD} measurements and $0.073^\circ$ for negative injections.
 
 The uncertainties in the catalog stellar positions of Trapezium stars referenced for our astrometric solution were taken from \cite{close2012}. These are a relative error on projected separations of 0.25\% and an absolute error in PA of $0.3^\circ$. The catalog astrometry of \cite{close2012} was itself referenced to Hubble Space Telescope (HST) observations of the Trapezium. The astrometric uncertainties for the underlying HST observations were taken from \cite{vandermarel2007} as a relative error of 0.0013\% in projected separation and an absolute error of $0.0028^\circ$ in PA. The absolute uncertainties on the DX-aperture detector-plane projected separation measurements due to their magnification to match the nominal SX platescale are $0.72$ pixels at either nod position and with either of the \texttt{GCNTRD} or negative injection techniques.
 
 The relative uncertainty in the distortion-corrected plate scale for the SX- and (magnified) DX-aperture images is $0.47\%$. The final source of uncertainty considered in this work is the uncertainty in true-north rotation angles for the distortion-corrected LBTI/LMIRCam images, which were computed separately for the SX and DX apertures of the telescope (see Section \ref{sssec:solns}). These uncertainties are $0.225^\circ$ in PA for the SX-aperture images and $0.254^\circ$ in PA for the DX-aperture images.
 
The detector-plane error sources on each measurement were added in quadrature, i.e., they were assumed to be independent. From the four astrometric measurements with either technique, mean values and standard errors on the mean were computed for the projected separation and PA of \textsc{Hii} 1348B. The standard error in each case was calculated from the inverse-variance weighted mean but with the variance of the mean inflated by a factor of 2 to account for the only semi-independence between the four data sets. This is analogous to computing the standard error on the mean as $\mathrm{SE} = \sigma / \sqrt{2}$ instead of $\mathrm{SE} = \sigma / \sqrt{N_\text{data sets} = 4}$ in the case of identical uncertainties represented by $\sigma$. We can therefore conclude that our errors on the mean of 5 mas in projected separation and $0.3^\circ$ in PA are likely over-estimated. We take the mean negative-injection relative astrometry with associated error as our reported companion position, given the slightly smaller errors compared with those yielded by the \texttt{GCNTRD} fitting procedure. Our relative astrometry of \textsc{Hii} 1348B is reported together with that of the four earlier observation epochs in Table \ref{tab:astrometry}.
 
 The largest source of uncertainty in the on-sky projected separation of \textsc{Hii} 1348B for the mean negative-injection relative astrometric measurement---reported in Table \ref{tab:astrometry}---is the uncertainty in plate scale used to convert pixel separation in the detector plane to angular separation on the sky. This conversion to on-sky coordinates contributes ${\sim}53\%$ of the variance in projected separation. Thus, improvements to the precision and/or accuracy of the plate scale solutions calculated for LBTI/LMIRCam have the most potential to significantly reduce such uncertainties in measurements of on-sky projected separation. The largest source of uncertainty in our reported PA of \textsc{Hii} 1348B is the fiducial uncertainty of the reference catalog astrometry for the Trapezium cluster published by \cite{close2012} of 0.3\textdegree, contributing ${\sim}59\%$ of the total variance in our reported PA.

\begin{figure*}
\plotone{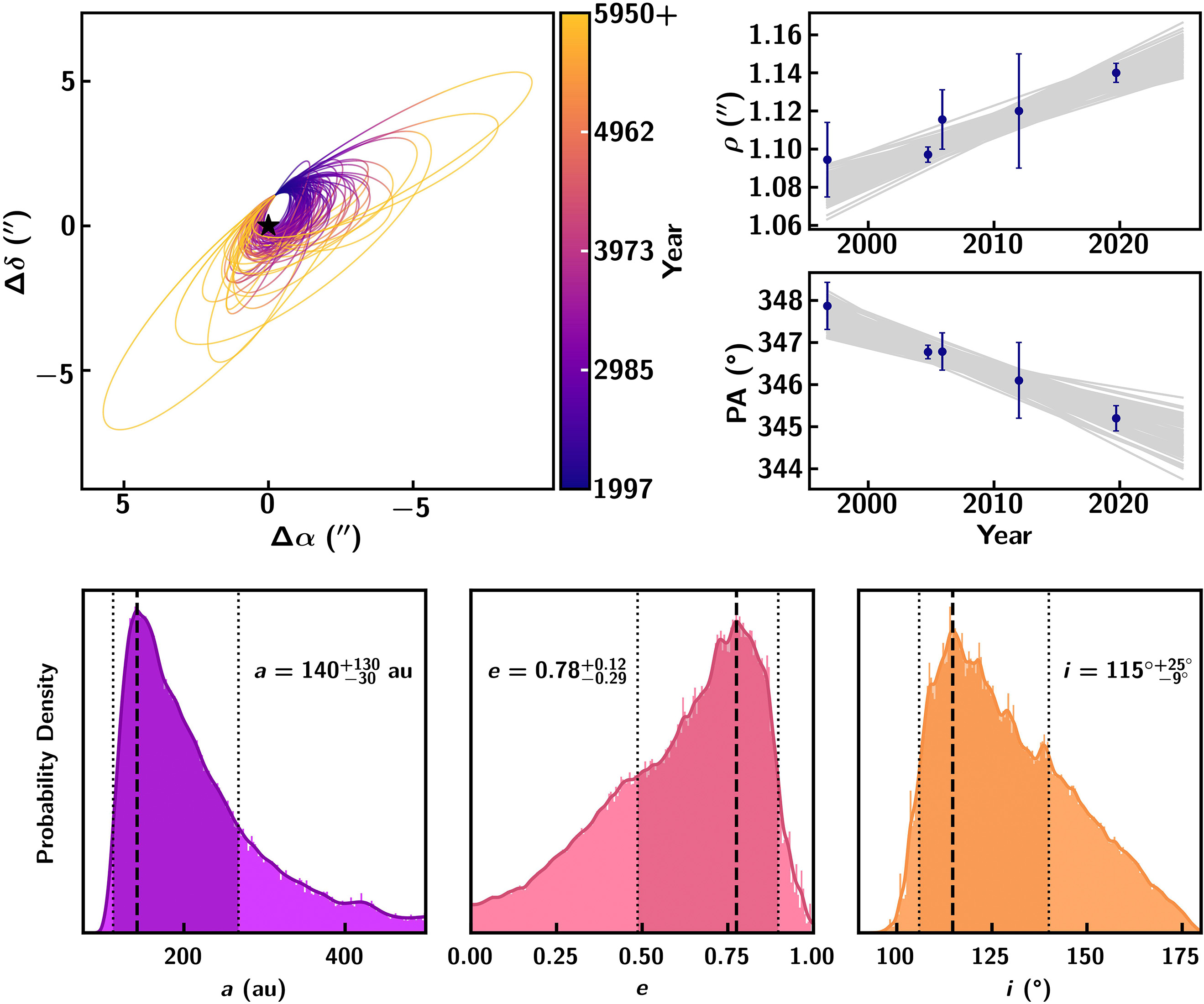}
\caption{Top left: 100 randomly selected posterior orbits of \textsc{Hii} 1348B from fitting to relative astrometry using \texttt{orbitize!}\ with a Markov Chain Monte Carlo (MCMC) sampler \citep{blunt2020}. The orbits are projected into relative right ascension and declination $(\Delta \alpha, \, \Delta \delta)$, with color indicating the year along the orbits. The lower limit of the colorbar corresponds to the first relative astrometric epoch from \cite{bouvier1997} and the upper limit is the mean epoch for the plotted orbits to complete their first orbit. Top right: in blue are the projected separation and position angle $(\rho, \, \text{PA})$ of \textsc{Hii} 1348B as a function of time. Grey tracks show the $\rho$ and PA of the same 100 posterior orbits plotted on the left. Bottom: histograms of selected orbital elements for all $10^7$ posterior orbits from the converged MCMC chains. Solid, colored lines show kernel density estimates of the marginalized distributions. Dashed and dotted black lines show the locations of the maximum a posteriori (MAP) estimates and boundaries of 68\% highest posterior density (HPD) regions, respectively.
\label{fig:orbits}}
\end{figure*}

\subsection{Orbit Fitting}\label{ssec:orbit_fitting}
Using our astrometry together with that reported in \cite{bouvier1997}, \cite{geißler2012}, and \cite{yamamoto2013} (see Table \ref{tab:astrometry}) we fit Keplerian orbits to the companion's projected motion. We generated a sample of $10^7$ posterior orbits of \textsc{Hii} 1348B with both the Orbits for the Impatient \citep[OFTI;][]{blunt2017} Bayesian rejection-sampling algorithm and a Markov Chain Monte Carlo (MCMC) algorithm implemented in the \texttt{orbitize!}\ Python package \citep{blunt2020}.

 The MCMC sampling algorithm used was the \texttt{ptemcee} \citep{vousden2016} adaptive parallel-tempered fork of the \texttt{emcee} \citep{foreman_mackey_2013} affine-invariant sampler. OFTI is a Bayesian rejection-sampling algorithm designed explicitly for imaged long-period companions with available relative astrometry covering only a small fraction of an orbit. Under certain conditions relevant to such wide-orbit companions, OFTI has been shown to converge to the same posteriors as MCMC samplers (within their shot noise) in significantly less time \citep{blunt2017}. Trial orbits are accepted or rejected within the OFTI algorithm based on comparing an orbit's likelihood to a uniform random number \citep{blunt2020}. OFTI also restricts the ranges of the input priors on inclination and system mass based on the results of the first 100 accepted orbits, inferring conservative upper and lower limits to place on these priors without otherwise changing their shape \citep{blunt2020}.

Stellar RV measurements for the host binary were not included in the orbital modeling due to the complications introduced by the binarity of the host stars and the expected very long orbital period of \textsc{Hii} 1348B (${\gtrsim}$ 1000 yr). Additionally, \textsc{Hii} 1348 was not included in the Hipparcos catalog \citep{hipparcos}, and thus astrometric accelerations from the Hipparcos--Gaia Catalog of Accelerations \citep{hgca_brandt_2021} are not available for inclusion into the orbital modeling of this substellar companion. This leaves determining a dynamical mass for \textsc{Hii} 1348B open to future studies.

We applied default priors in the \texttt{orbitize!}\ code to the companion's orbital elements. These are: uniform on $[0, 1]$ for eccentricity ($e$) and the dimensionless epoch of periastron passage ($\tau$), uniform on $[0, 2\pi]$ for the argument of periastron ($\omega$), uniform on $[0, \pi]$ for the position angle of nodes ($\Omega$), $\sin[0,\pi]$ for inclination ($i$), and log-uniform on $[10^{-3}, 10^4]$ au for semimajor axis ($a$). The uniform prior on eccentricity for \textsc{Hii} 1348B can be motivated by the estimated mass of this companion being near the substellar/stellar boundary and the observed eccentricity distribution of long-period stellar binaries being close to a uniform distribution \cite[e.g.,][]{duchene_kraus_2013}. The sine prior on $i$ and uniform priors on $\omega$ and $\Omega$ correspond to random orientations of orbits in space, restricted to one of two degenerate modes (see Section \ref{sssec:orbit_fitting_results} for further discussion of this degeneracy). Additionally, a Gaussian prior on the total mass of the system was applied as $M_\text{tot} = 1.28 \pm 0.09 \ \mathrm{M_\odot}$, the sum of the host binary mass estimated by \cite{geißler2012}: $1.22 \pm 0.09 \ \mathrm{M_\odot}$ and a mass estimate of $63 \pm 3 \ \mathrm{M_J}$ for the wide companion from comparison of its $L^\prime$ photometry to the evolutionary models of \cite{baraffe2003} (see Section \ref{ssec:mass} for more discussion of our substellar mass estimates from evolutionary models). A Gaussian prior was also applied for the system parallax, corresponding to the Gaia Early Data Release 3 \citep[EDR3;][]{gaiaedr3} parallax measurement for the system of $\pi = 6.979 \pm 0.0305 \ \text{mas}$. These Gaussian priors were applied such that nonphysical negative values of mass or parallax could not be drawn.

\subsubsection{MCMC Convergence}\label{sssec:mcmc_convergence}
 Before $10^7$ posterior orbits were drawn from the MCMC chains, $10^7$ samples were discarded as ``burn-in'' to encourage the convergence of the MCMC chains to sample the posterior distribution. With the MCMC parameters we supplied to the parallel-tempered scheme of the \texttt{ptemcee} sampler in \texttt{orbitize!}, this corresponded to the first 30,000 steps of 1,000 walkers being discarded, and the subsequent 30,000 steps being recorded, with a thinning of 3 (i.e., only every third step was recorded). This process yielded $10^7$ discarded burn-in samples and the $10^7$ converged samples for analysis. With parallel tempering, the 1,000 walkers generating the samples were only the 1,000 lowest-temperature walkers out of 20,000 total walkers distributed uniformly across 20 temperatures. The parallel-tempered algorithm of \texttt{ptemcee} periodically proposes ``swaps'' in the positions of walkers that are typically at adjacent temperatures, where each swap proposal will be accepted with a probability related to the relative likelihood of the two positions and the temperature difference \citep{vousden2016}. With this swapping, low-temperature walkers can efficiently and deeply explore widely separated modes (i.e., regions of the parameter space with high posterior probability density) by accessing positions of higher-temperature walkers which asymptotically sample the prior distribution with increasing temperature \citep{vousden2016}.

 The 20,000 MCMC walker initial positions were drawn from a posterior distribution yielded by fitting the orbit of \textsc{Hii} 1348B with the OFTI sampler. We expect this initialization of walker positions from the OFTI posterior to reduce the number of steps required for convergence of the MCMC chains by allowing them to very quickly begin exploring regions of parameter space with high inferred posterior probability density. By default, \texttt{orbitize!}\ would initialize MCMC walker positions to random draws from the prior distribution.

\begin{figure*}
\plotone{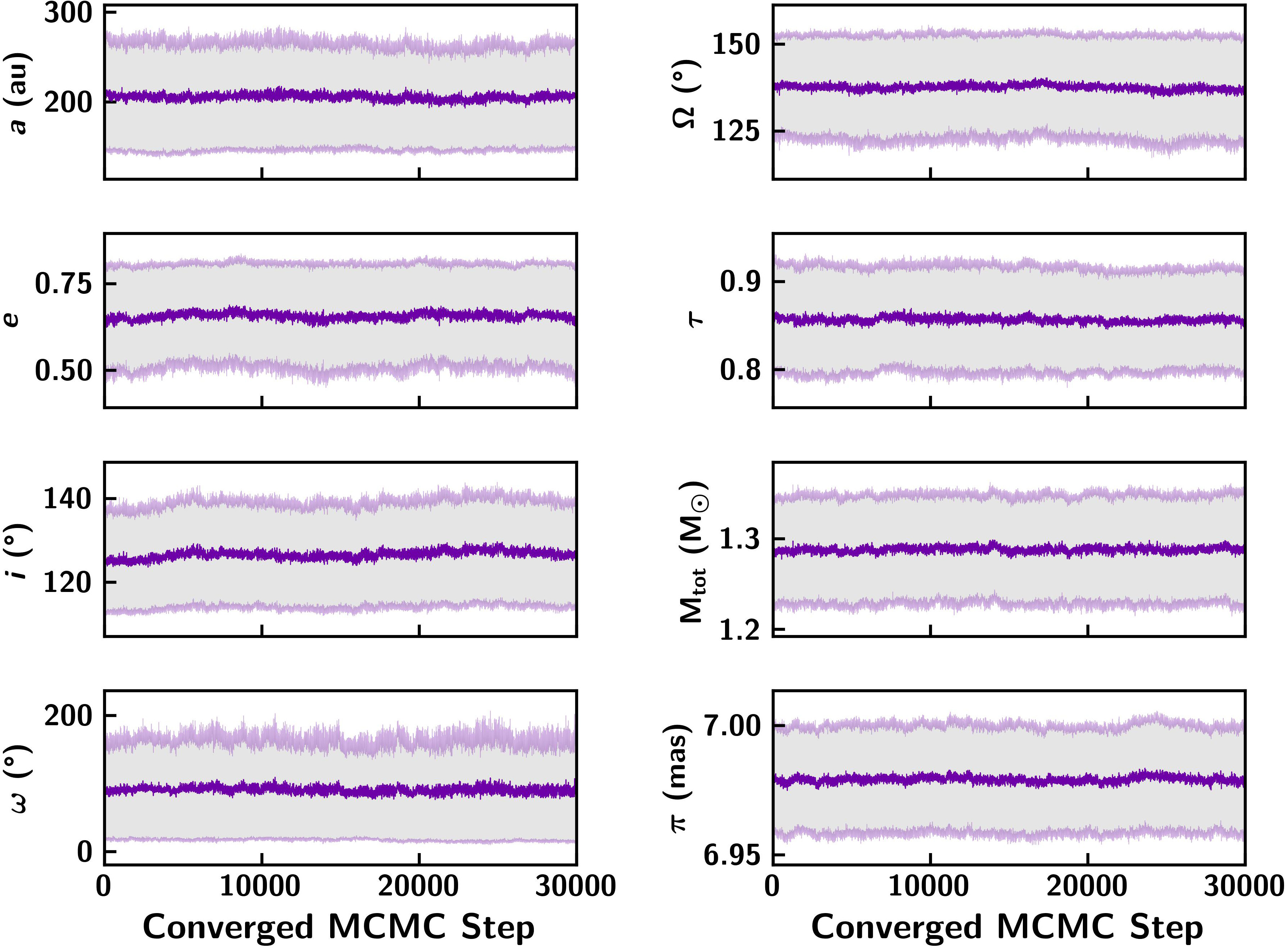}
\caption{ Median $\pm$ MAD lowest-temperature MCMC walker positions at each saved step in the converged MCMC chains. By visual inspection, the constant median and MAD values for each fitted parameter across 30,000 steps, along with clear distinctions from the priors, indicate that the chains have converged to the posterior distribution. Only every third sample was saved to reduce the correlation between steps (i.e., ``thinning''), giving $10^7$ samples for 1,000 walkers at the lowest temperature across $30\text{,}000 / 3 = 10\text{,}000$ thinned steps. The degeneracy in $\omega$ and $\Omega$ for fitting only to relative astrometric measurements means that these two parameters could equally likely both be offset by 180\textdegree $\pmod{360^\circ\text{, see Section \ref{sssec:orbit_fitting_results}}}$.
\label{fig:MCMC_chains}}
\end{figure*}

  The median and median absolute deviation (MAD) of the 1,000 lowest-temperature walkers at each recorded step after the burn-in period are plotted in figure \ref{fig:MCMC_chains}. Visual inspection of the approximate constancy of the median and MAD values of walker positions across the 30,000 post-burn-in steps suggests that the walker chains have converged to the posterior distribution. The median and MAD values across the post-burn-in steps are also clearly distinct from the assumed prior distributions (except for total system mass and parallax, which are consistent with the applied Gaussian priors), indicating that the relative astrometric data have contributed significant constraints to the posterior.

\subsubsection{Orbit-Fitting Results}\label{sssec:orbit_fitting_results}
 By studying the distributions of orbital elements yielded by fitting with \texttt{orbitize!}/MCMC, we draw conclusions about this brown dwarf companion's possible formation mechanisms. In Figure \ref{fig:orbits} we show 100 of the $10^7$ MCMC posterior orbits plotted along with their $\rho$ and PA as functions of time. Five epochs of relative astrometry across 23 years are included in our modeling (see Table \ref{tab:astrometry}). The precision of our new LBTI/LMIRCam $L^\prime$-band astrometric measurement with an uncertainty of ${\sim}5$ mas significantly constrains the probable orbits of \textsc{Hii} 1348B. Orbit fitting was also performed with the apparent outlier removed---the ``Nod 2'' SX-aperture astrometry---which yielded the lowest relative declination ($\Delta \delta$) for \textsc{Hii} 1348B of the four data sets (see Figure \ref{fig:quad_astro}). The exclusion of this outlying measurement did not significantly impact the resultant posterior distributions, with 1-D marginalized maximum-a-posteriori (MAP) estimates for orbital elements differing by ${\lesssim}3\%$ from those obtained by using the average astrometry of all four measurements.

 We find from the MCMC posterior distribution of fitted orbits that \textsc{Hii} 1348B has a semimajor axis and eccentricity of $a = 140 \substack{+130 \\ -30} \ \mathrm{au}$ and  $e = 0.78 \substack{+0.12 \\ -0.29}$, along with an inclination of $i = 115^\circ \substack{+ 25^\circ \\ -9^\circ}$. These and the other values for orbital elements that we report (see Table \ref{tab:system} and Figure \ref{fig:corner_plot}) inferred MAP estimates with uncertainties bounding the 68\% highest posterior density (HPD) credible regions of the appropriate 1-D marginalized distribution. HPD regions were computed in this study with \texttt{arviz.hdi} in Python \citep{arviz}. MAP estimation was performed via finding the peaks of kernel density estimates (KDEs) for each 1-D marginalized distribution. These KDEs and others computed in this study used the \texttt{KDEpy.FFTKDE}\footnote{\url{https://kdepy.readthedocs.io/en/latest/API.html\#fftkde}} Python code, based on the implementation of \texttt{statsmodels.nonparametric.kde.KDEUnivariate} \citep{statsmodels}. These KDEs, inferred MAP estimates, and 68\% HPD regions are over-plotted on the 200-bin histograms in Figures \ref{fig:orbits} and \ref{fig:corner_plot}, which include a representative sample of visual orbits and the full posterior distribution from fitting with \texttt{orbitize!}/MCMC. The posterior yielded with the OFTI sampler are consistent with the MCMC results (e.g., $a_\text{OFTI} = 150 \substack{+110 \\ -31} \ \mathrm{au}$, $e_\text{OFTI} = 0.78 \substack{+0.12 \\ -0.36}$, $i_\text{OFTI} = 116^\circ \substack{+26^\circ \\ -9^\circ}$), though with fewer spurious peaks in the posterior.

The likely high eccentricity of \textsc{Hii} 1348B's orbit can be contextualized within the results of recent works where directly imaged brown dwarf companions are found to generally have higher orbital eccentricities than directly imaged giant planets---hinting at a possible difference in formation channels for the two substellar populations \citep[e.g.,][]{bowler2020, nagpal2023, doo2023}. However, the 68\% HPD region on the orbit's eccentricity should be interpreted with some caution, as \cite{ferrer_chavez_2021} show that simulated direct-imaging observations covering a small fraction of the orbital period can lead to an eccentricity posterior that often under-covers the true value. \cite{ferrer_chavez_2021} show that the true eccentricities of orbits with similar phase coverage to that of \textsc{Hii} 1348B may have less than the expected 68\% probability of falling within such a credible region. For further discussion of the choices of priors for orbital elements and their possible biases, see Appendix \ref{appendix:A}.

The argument of periastron $\omega$ is constrained by our fitting modulo a $180^\circ$ degeneracy, with peaks in posterior probability density around $13^\circ$ and $193^\circ$. The position angle of nodes ($\Omega$) is also constrained only up to a $180^\circ$ degeneracy for the same cause, with peaks in its marginalized posterior around $141^\circ$ and $321^\circ$. This degenerate pair of posterior modes arises when fitting only to relative astrometric data where the sign of the companion's RV, i.e., motion into or out of the plane of the sky, is undetermined. We restricted $0 \leq \Omega \leq 180$ in our fitting to aid convergence of the MCMC walkers to one such mode, and subsequently apply the transformation $\omega^\prime = \omega + \pi$, $\Omega^\prime = \Omega - \pi$ to yield the other mode for reporting pairs of MAP estimates and credible regions. In \texttt{orbitize!}, $\tau$ is a dimensionless epoch of periastron, defined as $\tau = (t_\mathrm{p} - t_\mathrm{ref})/P_\mathrm{orb}$, where $t_\mathrm{p}$ is the epoch of periastron, $t_\mathrm{ref}$ is a reference time within one orbital period of $t_\mathrm{p}$ (so that $0 < \tau < 1$), and $P_\mathrm{orb}$ is the orbital period. We used the (arbitrary) \texttt{orbitize!}\ default value of 58849 MJD (2020 January 1) for $t_\mathrm{ref}$. The 1-D marginalized posterior distribution for $\tau$ is approximately Gaussian, corresponding to $\tau = 0.85 \substack{+0.11 \\ -0.07}$, with a spurious peak at $\tau = 1$ (i.e., periastron at the arbitrary reference epoch) that does not significantly bias the reported MAP estimate and HPD region. Such spurious peaks are present in some of the marginalized posterior distributions from the MCMC sampler, even for a large sample of $10^7$ converged walker positions. We found that the posterior yielded by the OFTI sampler was generally smoother for the same sample size, but we have chosen to highlight the MCMC sampling in this study given that it is the more well-established and well-vetted method for sampling from multi-dimensional posterior distributions in the literature.

\subsection{Photometry}\label{ssec:photometry}
In Section \ref{ssec:astrometry}, our method of negative artificial companion grid searches for determining the relative astrometry and contrast of \textsc{Hii} 1348B was discussed. This process, with cADI star-subtraction, yielded four residual-minimizing companion contrasts which were combined through a mean. Similar to the astrometric procedure (see Section \ref{ssec:astrometry}), the error on the mean contrast ratio for \textsc{Hii} 1348B was computed as the inverse-variance weighted mean, with the variance of the mean inflated by a factor of two to account for the only semi-independence of the four data sets split by aperture and nodding position. The uncertainty on each of the four contrast ratios was taken as the sample standard deviation of the measurements. Sources of uncertainty in the host star's luminosity are only relevant when we infer an apparent magnitude of \textsc{Hii} 1348B from the measured contrast ratio. We report a mean companion contrast ratio (\textsc{Hii} 1348B with respect to the unresolved host binary) of $(9.4 \pm 0.2) \times 10^{-3}$ in the $L^\prime$ filter. This contrast ratio is plotted above our $5\text{-}\sigma$ detection limits determined from synthetic injections and retrievals (see Section \ref{ssec:det_limits}) in Figure \ref{fig:contrast_curve}.

To compute the $L^\prime$-band magnitude of \textsc{Hii} 1348B corresponding to this measured contrast ratio, we used Wide-field Infrared Survey Explorer All-Sky Data Release \citep[AllWISE:][]{wright2010, cutri2014} and Two Micron All Sky Survey \citep[2MASS:][]{2MASS, cutri2003} reported photometry for \textsc{Hii} 1348A. The $W1$-band ($\lambda_\mathrm{c} = 3.466 \ \mu\mathrm{m}$, $\Delta \lambda_\text{FWHM} = 0.636  \ \mu\mathrm{m}$) inner binary host star magnitude of $9.615 \pm 0.023$ was taken together with an additional relative error to account for variability. \textsc{Hii} 1348A is variable in the $V$ bandpass at ${\sim}5\%$ in flux, as measured by the All-Sky Automated Survey for SuperNovae \citep[ASAS-SN:][]{jayasinghe2019} catalog of variable stars. We expect the variability to be somewhat less in the mid-IR than in the optical $V$ bandpass but conservatively adopt the same 5\% uncertainty in flux. We then estimated a color correction from $W1$ to $L^\prime$ by fitting a quadratic to the 2MASS $K_S$ ($\lambda_\mathrm{c} = 2.163 \ \mu\mathrm{m}$, $\Delta \lambda_\text{FWHM} = 0.278 \ \mu\mathrm{m}$), $W1$ ($\lambda_\mathrm{c} = 3.466 \ \mu\mathrm{m}$, $\Delta \lambda_\text{FWHM} = 0.636  \ \mu\mathrm{m}$), and $W2$ ($\lambda_\mathrm{c} = 4.644 \ \mu\mathrm{m}$, $\Delta \lambda_\text{FWHM} = 1.107 \ \mu\mathrm{m}$) spectral flux densities of \textsc{Hii} 1348A. We interpolated to $L^\prime$ nominally at $\lambda_\mathrm{c} = 3.700 \ \mu\mathrm{m}$ to estimate $W1 - L^\prime \approx 0.033$. Accounting for variability and applying the aforementioned color correction, we adopt the unresolved host binary to have an apparent magnitude of $L^\prime_\text{A} = 9.58 \pm 0.06$. The mean contrast ratio of \textsc{Hii 1348B} to \textsc{Hii 1348A} in our $L^\prime$-band imaging was then multiplied by the host star's spectral flux density at this magnitude to compute our reported $L^\prime_\text{B} = 14.65 \pm 0.06$ for the companion. In Section \ref{ssec:mass}, we describe mass estimates for the companion using bolometric luminosities inferred from this photometry independently as well as included in an average across all available photometry of \textsc{Hii} 1348B.

\subsection{Spectral Fitting}
\label{ssec:spectral_fitting}
\begin{figure*}
\plotone{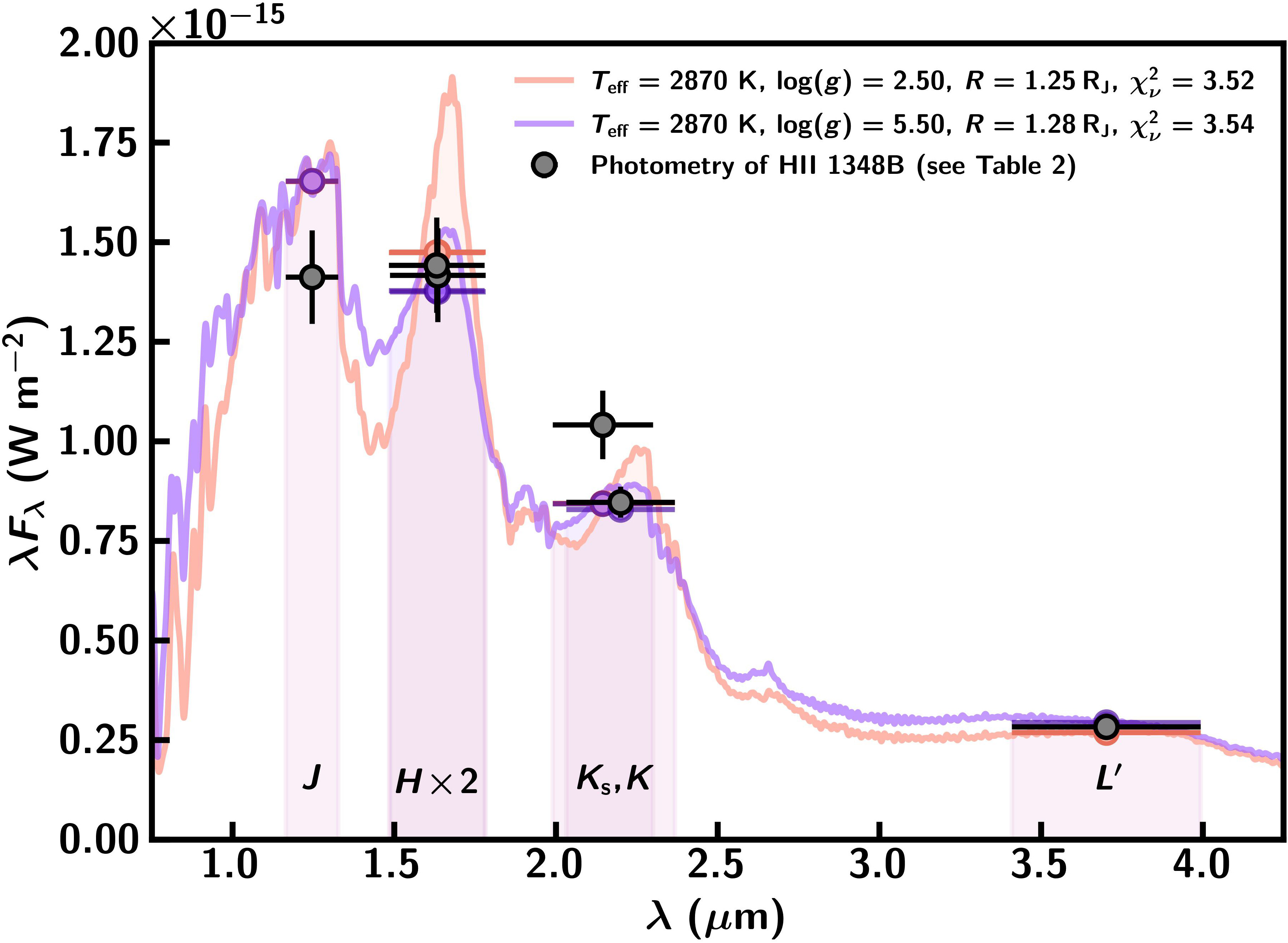}
\caption{ Measured apparent photometry of \textsc{Hii} 1348B from Table \ref{tab:system} is plotted in black with interpolated \textsc{bt-settl} (CIFIST 2011/2015) model spectra from MAP estimation plotted in salmon and purple. Synthetic photometry as integrated from the model spectra are shown in their respective colors with horizontal error bars spanning the shaded filter widths ($\Delta \lambda_\text{FWHM}$). Nonphysical, very low and very high $\log(g)$ at grid edges are yielded by MAP estimation of the two posterior modes. The multi-wavelength photometry of \textsc{Hii} 1348B is inferred to be generally consistent with a late-M spectral type, with no constraint on $\log(g)$. The interpolated model spectra are smoothed with a Gaussian filter to an average spectral resolution of $R = 250$ for the displayed wavelengths. Reduced chi-square values ($\chi_\nu$) are computed for three degrees of freedom: $T_\text{eff}$, $\log(g)$, and $R$.
\label{fig:spectrum}}
\end{figure*}

 This work provides the first $L^\prime$-band photometry of \textsc{Hii} 1348B and we have fit \textsc{bt-settl} Cosmological Impact of the First Stars (CIFIST) 2011/2015\footnote{\url{http://svo2.cab.inta-csic.es/theory/newov2/index.php?models=bt-settl-cifist}}effective temperature and surface gravity of \textsc{Hii} 1348B. Without including a direct spectrum of \textsc{Hii} 1348B in this analysis, we primarily seek to check for consistency of our $L^\prime$-band photometry with the $\mathrm{M}8 \pm 1$ spectral type determined by \cite{geißler2012}. The six available photometric measurements of \textsc{Hii} 1348B, spanning ${\sim}1.2\text{--}3.7 \ \mu \mathrm{m}$, and two consistent model spectra are shown in Figure \ref{fig:spectrum}. The Keck/OSIRIS and Palomar/PHARO spectra of \textsc{Hii} 1348B presented in Figures 4 and 5 of \cite{geißler2012} are not readily available, and we did not make an effort to reproduce them (e.g., from digitizing the published plots).

 The spectra plotted in Figure \ref{fig:spectrum} are the result of interpolation between adjacent \textsc{bt-settl} (CIFIST 2011/2015) model grid points. These models are fixed at solar abundances\footnote{See Appendix \ref{appendix:B} for a discussion of our initial fitting to a larger \textsc{bt-settl} grid varying over metallicity and alpha enhancement.} determined by \cite{caffau_2011}. \textsc{bt-settl} atmospheric models incorporate cloud and dust formation, making them well suited for late-M and L-type dwarfs with cloudy photospheres \citep{allard2012, allard2013}. In addition, the CIFIST 2011/2015 models considered in this work include refinements not present in other \textsc{bt-settl} grids, including updated molecular line lists and calibrated atmospheric convection parameters \citep{baraffe_2015}. The grid spans surface gravities $2.5 \leq \log(g \ \text{cm}^{-1} \ \text{s}^{2}) \leq 5.5$ (hereafter ``$\log(g)$'') in intervals of 0.5 dex, and effective temperatures $1200 \ \mathrm{K} \leq T_\text{eff} \leq 7000 \ \mathrm{K}$ in intervals of either 50 or 100 K. We have restricted our search to $2000\text{--}4100 \ \mathrm{K}$ and omitted the $50 \ \mathrm{K}$--interval models to create a uniform grid of 154 models separated by intervals of 100 K in $T_\text{eff}$ and 0.5 dex in $\log(g)$. To help fill the relevant parameter space, an additional 51 model atmospheres were created between each pair of adjacent models in the grid via linear interpolation of $\log(F_\lambda)$ over $g$ and $T_\text{eff}$ (see the ``Model Grid'' in Figure \ref{fig:spectrum_post_surface}).

 The model spectra are computed per unit area on the object's surface, so we have scaled the models by dilution factors $\Omega = R^2/d^2$, where $R$ is the inferred radius of the brown dwarf and $d$ is the physical distance to the system. For \textsc{Hii} 1348, $d = (143.3 \pm 0.6) \ \mathrm{pc}$ \citep{gaiaedr3}. $\Omega$ is optimized to minimize the chi-square ($\chi^2$) value for each fitted model. $\chi^2$ was computed as,

 \begin{equation}
    \chi^2 = \sum_i (F_{i} - F_{\text{syn},i})^2/\sigma_{i}^2
\end{equation}

 with the sum over $i$ taken over the six photometric fluxes of \textsc{Hii} 1348B ($F_i$) with assumed symmetric Gaussian errors ($\sigma_i$) and synthetic photometry ($F_{\text{syn},i}$) computed from direct integration of the model spectra over the relevant bandpass. Synthetic photometry was calculated assuming simple rectangular transmission profiles for all filters, centered on the central wavelength and spanning the FWHM of the transmission curve.

\subsubsection{Spectral Fitting Results}\label{sssec:spec_fit_results}
The $K_S$-band photometry at $\lambda_\text{c} = 2.145 \ \mu\mathrm{m}$ was generally found to be the most discrepant measurement with the fitted models, contributing about half of the reduced chi-square ($\chi_\nu^2$) for the spectra inferred by MAP estimation shown in Figure \ref{fig:spectrum}.  With six photometric points and 3 degrees of freedom ($T_\text{eff}$, $\log(g)$, $R$), $\chi_\nu^2$ is minimized at 3.16 for the maximum-likelihood model, which lies at one edge of the model grid with $T_\text{eff} = 3100 \ \mathrm{K}$ and $\log(g) = 5.5$ dex. This model has a significantly higher effective temperature than expected for an object with a spectral type of ${\sim}$M8, and the very high surface gravity would imply an unrealistic mass for \textsc{Hii} 1348B of ${\sim} 170 \ \mathrm{M_J}$---well into the stellar regime.

\begin{figure*}
\plotone{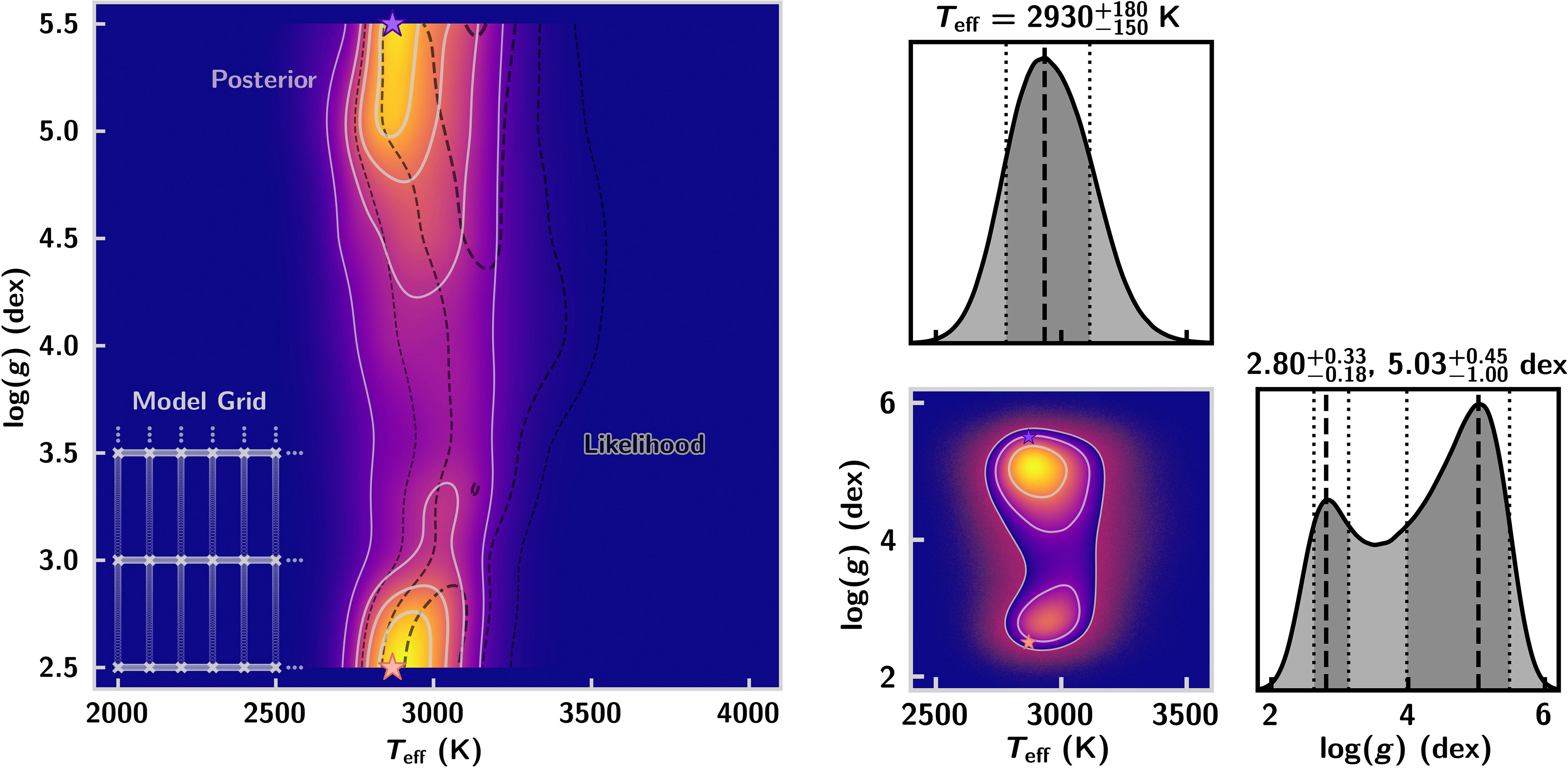}
\caption{ Left: inferred posterior surface for the tested grid of \textsc{bt-settl} (CIFIST 2011/2014) models fit to photometry of \textsc{Hii} 1348B assuming a Gaussian prior $T_\text{eff} = 2500 \pm 300 \ \mathrm{K}$ for a spectral type of $\mathrm{M8 \pm 1}$. The colormap represents relative inferred posterior probability density. A cut-out of the grid spacing is shown with light-gray ``x'' markers, and locations of fitted models interpolated between grid points are shown with open circles. The likelihood surface (equivalent to taking a uniform prior on $T_\text{eff}$) is shown with dark-gray dashed contours. Purple and salmon ``$\star$'' markers show the locations of inferred local posterior maxima for tested models (shown fitted to the observed photometry of \textsc{Hii} 1348B in Figure \ref{fig:spectrum}). Right: corner plot of inferred $T_\text{eff}$ and $\log(g)$ for $6.6 \times 10^6$ samples with added Gaussian noise drawn from the interpolated posterior with \texttt{emcee} \citep{foreman_mackey_2013}. Diagonal subplots show 200-bin histograms with KDEs over-plotted in solid black. Dashed and dotted lines mark MAP estimates and 68\% HPD regions for each parameter. The off-diagonal subplot shows inferred density with a 2-D KDE and scattered samples below the lowest contour.
\label{fig:spectrum_post_surface}}
\end{figure*}

 To incorporate our exceptions of $T_\text{eff}$ for \textsc{Hii} 1348B from its $\mathrm{M8} \pm 1$ spectral type \citep{geißler2012}, we compute posterior probabilities of fitted models with a Gaussian prior on effective temperature corresponding to $T_\text{eff} = 2500 \pm 300 \ \mathrm{K}$. This prior was computed from the polynomial relation for $T_\text{eff}$ as a function of IR spectral type published in Table 5 of \cite{sanghi_2023} for young ultracool dwarfs (${\lesssim}3000$ K), where we assumed a Gaussian distribution in the spectral type with a mean of M8 and a standard deviation of unity.  The polynomial relations of \cite{sanghi_2023} were derived for a sample of 189 isolated young objects of spectral types M6--T9 via direct integration of flux-calibrated spectra and photometry from the optical to mid-IR. The rms scatter around this polynomial relation published by \cite{sanghi_2023} is added in quadrature with the uncertainty propagated from the uncertainty in spectral type of $\pm 1$ for \textsc{Hii} 1348B \citep{geißler2012}.  The likelihood and posterior surfaces are interpolated between tested models in Figure \ref{fig:spectrum_post_surface} with radial basis function (RBF) interpolation in Python using \texttt{scipy.interpolate.RBFInterpolator} \citep{scipy} with a cubic kernel.

 After the application of a prior on $T_\text{eff}$, the masses inferred from the fitted $R$ and $\log(g)$ of the two models shown in Figure \ref{fig:spectrum} yielded by MAP estimation are $\approx 200$ and $0.2$ $\mathrm{M_J}$. Such masses are nonphysical for the well-constrained age of the Pleiades open cluster and the intrinsic brightness of \textsc{Hii} 1348B. In Section \ref{ssec:mass}, we discuss the masses implied by evolutionary models, which lie at ${\sim}60\text{--}63 \ \mathrm{M_J}$. The MAP-estimated spectral models having effective temperatures of between 2800 K and 2900 K would imply $T_\text{eff}$ for \textsc{Hii} 1348B that is consistent with some M7.5-type objects \citep[e.g., LHS 2645 and 2MASSW J2258066+154416;][]{testi_2009}.

If we choose to restrict the \textsc{bt-setll} (CIFIST 2011/2015) model grid to only those which imply $1 \ \mathrm{M_J} < M_\text{B} < 100 \ \mathrm{M_J}$ for their fitted $R$ and $\log(g)$, i.e., $M_\text{B} = 10^{\log(g)} R^2 / G$, then the low-$\log(g)$ mode seen in Figure \ref{fig:spectrum_post_surface} is excluded for implying masses below $1 \ \mathrm{M_J}$. At the high-$\log(g)$ mode, only models near the grid edges are excluded, with the mode at $\log(g) \sim 5.0$ dex retained. With this relatively conservative---but yet arbitrary---restriction to masses between 1 and 100 $\mathrm{M_J}$, the interpolated \textsc{bt-settl} (CIFIST 2011/2015) model yielded by MAP estimation has $T_\text{eff} = 2850 K$, $\log(g) = 5.0$ dex, and $R = 1.29 \ \mathrm{R_J}$---with $\chi_\nu^2 = 3.65$. This atmospheric model implies a mass of ${\sim}67 \ \mathrm{M_J}$, a value which is generally consistent with the predictions found from evolutionary models (see Section \ref{ssec:mass}).

 The right panel of Figure \ref{fig:spectrum_post_surface} shows a corner plot for $6.6 \times 10^6$ posterior samples drawn from the interpolated posterior surface on the left of the figure with \texttt{emcee} in Python \citep{foreman_mackey_2013}. Additional broadening has been added to the samples to account for the uncertainty in interpolation between grid points. This additional broadening was added as Gaussian noise to the samples with mean zero and standard deviation equal to half the grid spacing (i.e., 50 K in effective temperature and 0.25 dex in surface gravity) in either dimension. A consequence of adding this broadening to the samples is that it artificially shifts the local maxima in the posterior PDF away from the grid edges (purple and salmon markers are carried over to the right panel in Figure \ref{fig:spectrum_post_surface} from the parent surface shown at the left). The 1-D marginalized distributions for each parameter from these samples and 68\% HPD regions are shown in diagonal subplots of Figure \ref{fig:spectrum_post_surface}. While the distribution in $\log(g)$ is not well constrained, we may estimate the effective temperature of \textsc{Hii} 1348B as $2930 \substack{+180 \\ -150}$ K. In the off-diagonal subplot, three light-gray contours plot mark the 2-D Gaussian 1.5-, 1.0-, and 0.5-$\sigma$ levels (i.e, containing ${\sim}67.5$\%, ${\sim}39.3$\%, and ${\sim}11.8$\% posterior probability) inferred from a 2-D KDE. Posterior samples are scattered in this subplot below the lowest such contour.

 There is no strong evidence from the broadband photometry of \textsc{Hii} 1348B alone for any of these three models (the unrestricted local posterior maxima at grid edges shown in Figures \ref{fig:spectrum}, \ref{fig:spectrum_post_surface} and the restricted-mass MAP model implying $M \sim 67 \ \mathrm{M_J}$) over another. The likelihood ratio of the global posterior-maximizing model with $\log(g) = 2.5$ dex (salmon in Figure \ref{fig:spectrum}) to the best restricted-mass model is $\approx 1.22$. Similarly, the likelihood ratio of the MAP-estimated model of the second mode with $\log(g) = 5.5$ dex (purple in Figure \ref{fig:spectrum}) to the restricted-mass posterior-maximizing model is $\approx 1.18$. Trivially, the likelihood ratio of the two MAP-estimated models at the grid edges is then $\approx 1.03$. The very close (or identical) $T_\text{eff}$ values inferred for these models give very close (or identical) prior probabilities, and therefore also produce similar ratios of posterior density to the likelihood ratios. Thus, while the MAP-selected atmospheric models from the full \textsc{bt-settl} (CIFIST 2011/2015) interpolated grid tested in this study imply unrealistic masses for \textsc{Hii} 1348B, they are not significantly preferred over those which would na\"ively imply masses consistent with the predictions of evolutionary models (see Section \ref{ssec:mass} for a discussion of the masses inferred from evolutionary models of brown dwarfs). It is reasonable to infer from visual inspection of the atmospheric models presented in Figure \ref{fig:spectrum} that even a low-resolution spectrum of \textsc{Hii} 1348B could serve to constrain its surface gravity. Particularly, in the near-IR from the red end of the $J$ bandpass through the $H$-band filter (${\sim}1.4\text{--}1.7 \ \mu\mathrm{m}$), there is expected to be a very noticeable (i.e., measurable) gravity-induced suppression of spectral features at equal $T_\text{eff}$. Such constraining spectroscopy could allow for a stronger measure of the discrepancies between the bulk atmospheric properties of \textsc{Hii}1348B inferred from atmospheric and evolutionary models.

 In addition to the surface gravity of \textsc{Hii} 1348B remaining yet unconstrained, radius estimates from spectroscopic fitting can be subject to systematic effects and results are often discrepant with expectations from evolutionary models \citep[e.g.,][]{dupuy_kraus_2013, zhang_2021, sanghi_2023}. Our typical uncertainties in $R$ for a given dilution factor $\Omega$, coming only from the 0.6-pc uncertainty in distance to the system, are therefore significantly underestimated at ${\sim} 0.005 \ \mathrm{R_J}$. Where $\log(g)$ may be better constrained to infer mass, we refer the reader to the analysis of deriving radii with \textsc{bt-settl} (CIFIST 2011/2015) models presented by \cite{sanghi_2023}. Due to both the unconstrained surface gravity and the possibly unreliable radius estimates from fitting model spectra to photometry of \textsc{Hii 1348B}, we consider the mass estimates derived from evolutionary models in Section \ref{ssec:mass} to be much more robust than any such inferences on mass made from the \textsc{bt-settl} (CIFIST 2011/2015) models. Additionally, the masses of ultracool dwarfs predicted by evolutionary models typically agree with dynamical mass measurements \citep[e.g.,][]{dupuy_liu_2017, brandt2019, brandt2021}. See Section \ref{ssec:spec_fit_discussion} for further discussion of the parameters implied by fitting to the \textsc{bt-settl} (CIFIST 2011/2015) model grid, as compared with the results found for a large sample of young isolated late-M and L dwarfs by \cite{hurt_2024}.

\subsection{Detection Limits}\label{ssec:det_limits}

\begin{figure*}
\plotone{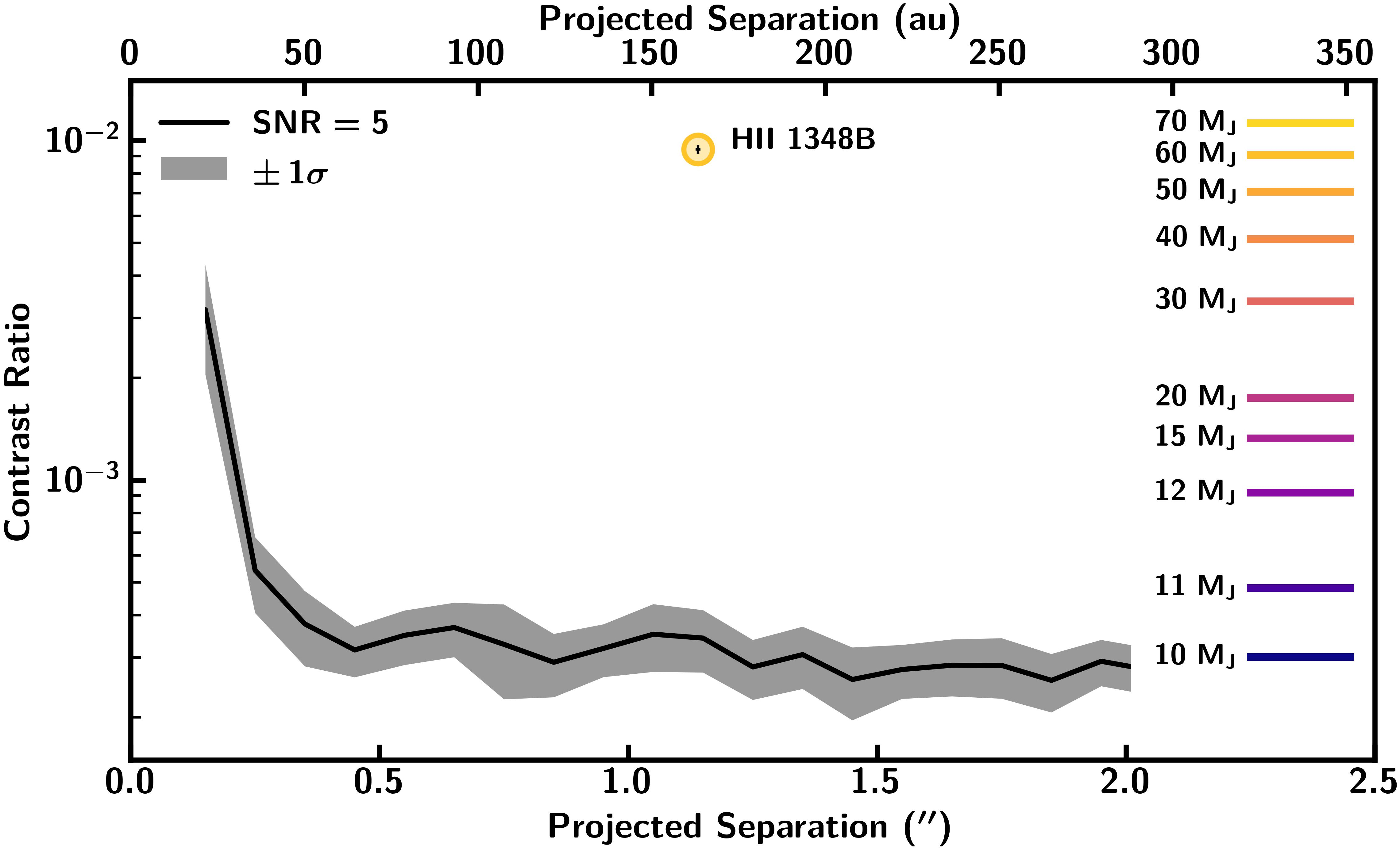}
\caption{Contrast curve for the sensitivity of our $L^\prime$ observations to companions as a function of projected separation. A nominal distance to the \textsc{Hii} 1348 system of $143.3 \ \text{pc}$ \citep{gaiaedr3} is assumed for the conversion of angular projected separation to au. In black is the mean contrast ratio required for retrieval within 5\% of $\mathrm{SNR} = 5$, while the shaded region shows the sample standard deviation across injections at 18 PAs. The observed $L^\prime$ contrast ratio and projected separation of \textsc{Hii} 1348B on 2019 September 18 are plotted in orange, with the error bars (smaller than the marker) over-plotted in black. Colored bars indicate expected $L^\prime$-band contrast ratios for substellar of varying masses at the assumed age of the Pleiades \citep[$112 \pm 5$ Myr;][]{dahm2015} generated from the evolutionary models of \cite{baraffe2003}.
\label{fig:contrast_curve}}
\end{figure*}

In Figure \ref{fig:contrast_curve}, the mean contrast ratio of point sources injected into our observations across 18 position angles for retrievals within 5\% of $\mathrm{SNR} = 5$ as a function of projected separation is shown in black. The standard deviation in injected contrasts yielding this SNR threshold across PAs is shaded in grey. The mass scale on the right of the figure was generated with the evolutionary models of \cite{baraffe2003} at the age of the Pleiades and distance to \textsc{Hii} 1348 assumed in this study (see Table \ref{tab:system}). We achieve detection limits down to $10\text{–}11 \ \mathrm{M_J}$ in the background limit at projected separations larger than ${\sim}0\farcs{25} \approx 36$ au, out to ${\sim}2'' \approx 290$ au. At the closest angular separations probed (${\sim}0\farcs{15} \approx 21.5$ au), we are sensitive only to higher-mass brown dwarfs with $M \gtrsim 30 \ \mathrm{M_J}$. No sources other than \textsc{Hii} 1348A and B were detected in our observations (see Figure \ref{fig:klip}).

\subsection{The Mass of \textsc{Hii} 1348B}\label{ssec:mass}
\textsc{Hii} 1348B has previously had its mass estimated to lie between 0.053 and 0.063 $\mathrm{M_\odot}$ \citep{geißler2012}. \cite{geißler2012} used evolutionary models from \cite{burrows1997} and \cite{chabrier2000}, bolometric corrections for M8 dwarfs \citep{dahn2002, golimowski2004, leggett2002}, an age of the Pleiades of $100\text{–}125 \ \mathrm{Myr}$, and  a distance to the Pleiades of either $120.2 \pm 1.9$ pc \citep{van_leeuwen_2009} or 133 pc---a weighted average of parallax distances from \cite{pan_2004}, \cite{munari_2004}, \cite{zwahlen_2004}, and \cite{southworth_2005}---to estimate the mass of \textsc{Hii 1348B} from $J$, $H$, and $K_S$ photometry. No uncertainty on the weighted-average parallax was reported by \cite{geißler2012}, so it is not clear if such a parallax uncertainty was considered in their computation of the absolute magnitude of \textsc{Hii} 1348B assuming this distance.

From the SEEDS survey \citep{SEEDS2009, SEEDS2014, SEEDS2016, SEEDS2017} High‐Contrast Instrument with Adaptive Optics \citep[HiCIAO:][]{HiCIAO} $H$-band observations with the Subaru telescope, \cite{yamamoto2013} reported a mass of 48 $\mathrm{M_J}$ for \textsc{Hii} 1348B. This result was referenced to the models of \cite{baraffe2003}, and \cite{yamamoto2013} took the Pleiades to have an age of 125 Myr at a distance of 135 pc.

 In this work, we have used the evolutionary models of \cite{burrows_2001} and \cite{baraffe2003} to estimate the mass of \textsc{Hii 1348B} from its bolometric luminosity ($L_\text{bol}$). $L_\text{bol}$ was estimated for each of the six photometric measurements shown in Table \ref{tab:system} via the empirical polynomial relations for young ultracool dwarfs as functions of absolute magnitudes published in Table 5 of \cite{sanghi_2023}. For each polynomial relation applied, the rms scatter of the polynomial fit given in Table 5 of \cite{sanghi_2023} was added in quadrature with the relevant propagated photometric uncertainty to derive Gaussian uncertainties in $L_\text{bol}$.

 We calculated absolute magnitudes of \textsc{Hii} 1348B in each of the six photometric bandpasses by assuming the Gaia-EDR3 distance measurement to the system of $143.3 \pm 0.6$ pc \citep{gaiaedr3}. A polynomial relation is not provided by \cite{sanghi_2023} for the $L^\prime$ filter. Instead, we estimated a contrast ratio correction factor from $L^\prime$ ($\lambda_\mathrm{c} = 3.70 \ \mu\mathrm{m}$, $\Delta \lambda_\text{FWHM} = 0.58 \ \mu\mathrm{m}$) to the WISE $W1$ bandpass ($\lambda_\mathrm{c} = 3.466 \  \mu\mathrm{m}$, $\Delta \lambda_\text{FWHM} = 0.636 \  \mu\mathrm{m}$) via integration of blackbody spectra across the two bandpasses at estimated effective temperatures of the system components: 4700 K and 4050 K for \textsc{Hii} 1348Aa and Ab, respectively. We estimate the contrast ratio in the $W1$ bandpass to be ${\sim}96\%$ of that measured in $L^\prime$ with LMIRCam, assuming rectangular transmission curves centered at $\lambda_c$ with a filter width $\Delta \lambda_\text{FWHM}$ in both cases.

 Averaging the six independent $L_\text{bol}$ estimates, we find $L_\text{bol} = -3.05 \pm 0.04$ dex. This value is referenced to the solar luminosity, i.e., $10^{-3.05 \pm 0.04} \ \mathrm{L_\odot}$, where $\mathrm{L_\odot} = 3.828 \times 10^{26}$ W. From this average value, we assumed a uniform prior on $L_\text{bol}$ between -3.09 and -3.01 dex and a Gaussian prior on the cluster/system age of $112 \pm 5$ Myr \citep{dahm2015} for comparison to evolutionary models. From the models of \cite{burrows_2001}, \textsc{Hii} 1348B is estimated to have a mass of $60 \pm 2 \ \mathrm{M_J}$. In agreement with this prediction within $1.06 \sigma$, we obtain from the models of \cite{baraffe2003} a mass estimate of $63 \pm 2 \ \mathrm{M_J}$. These values are also both in excellent agreement with the estimates published by \cite{geißler2012}, who inferred a range of masses between 55 and 66 $\mathrm{M_J}$ via comparison of absolute and bolometric magnitudes to the evolutionary models of \cite{chabrier2000} and \cite{burrows1997}.

As a check for consistency of the $L^\prime$ photometry measured in this study with the other photometric measurements, we also estimated the mass of \textsc{Hii} 1348B via comparison to the \cite{baraffe2003} models from our measurement of $L_\text{B}^\prime = 14.65 \pm 0.06$ alone. This yielded a posterior mass estimate of $63 \pm 3 \ \mathrm{M_J}$, consistent with the values obtained from the mean inferred bolometric luminosity of \textsc{Hii 1348}B. This value was taken as the companion's contribution to the total system mass prior used in our orbit fitting with \texttt{orbitize!}\ (see Section \ref{ssec:orbit_fitting}). Our mass estimate from $L^\prime$ photometry provides an independent confirmation with the first mid-IR imaging of the system that  \textsc{Hii} 1348B is highly likely to be substellar (${\sim}99\%$ probability) at the age and distance of the Pleiades, assuming a hydrogen-burning limit of ${\sim}70 \ \mathrm{M_J}$ \citep[e.g.,][]{dupuy_liu_2017, morley_2024}.

The discrepancy in mass estimates from this work and that of \cite{yamamoto2013} may be explained by our use of a larger distance to \textsc{Hii} 1348 as measured by Gaia \citep[$143.3 \pm 0.6 \ \mathrm{pc}$;][]{gaiaedr3}. This measurement was unavailable to \cite{yamamoto2013} at the time of their publication. Their assumed shorter distance to the system of 135 pc would imply a lesser intrinsic brightness---and therefore a smaller mass---of \textsc{Hii} 1348B at the same cluster age. However, the age of the Pleiades assumed by \cite{yamamoto2013} is nominally 13 Myr older than the $112 \pm 5 \ \mathrm{Myr}$ \citep{dahm2015} assumed in this study, which would produce the opposite bias towards predicting a slighter larger mass than our estimate at the same distance to the system. However, this is expected to have a smaller effect on the discrepancy in mass estimates than that from the difference in assumed distances given that only minimal cooling is expected to occur over this 13 Myr \citep[e.g.,][]{baraffe2003}. A measurement of the dynamical mass of \textsc{Hii 1348B} would help to confirm or reject these estimates which are inherently tied to relatively uncertain models of substellar evolution. Such dynamical measurements \citep[e.g.,][]{crepp2012, brandt2019, brandt2021} allow for the masses of giant-planet and brown-dwarf companions to be derived independently of any atmospheric or evolutionary models.

\subsection{Binary Host Star}
The double-lined spectroscopic binary (SB2) star \textsc{Hii} 1348A is composed of two K-type dwarfs with masses estimated as $0.67 \pm 0.07 \ \mathrm{M_\odot}$ (\textsc{Hii} 1348Aa) and $0.55 \pm 0.05 \ \mathrm{M_\odot}$ (\textsc{Hii} 1348Ab) \citep{geißler2012}. These masses were inferred by \cite{geißler2012} from the $B-V$ colors of each binary component reported by \citep{queloz1998}.  \cite{queloz1998} found the $B-V$ colors of the two SB2 binary components by measuring the rotational broadening of each of the stars' resolved spectral lines and inferring a mass ratio from the relative contrast of their cross-correlation function. Such close stellar binaries are not uncommon in the Pleiades, where the close-binary fraction (CBF; $a < 10 \ \mathrm{au}$) is $25\% \pm 3\%$ \citep{torres2021}. However, triple and higher-order configurations are generally uncommon for main-sequence (MS) and non-embedded pre-MS star systems with K-type primaries---the triple/high-order fraction (THF) for such systems is ${\sim}5\%$ \citep{offner2023}. The THF has been found to increase monotonically with mass/lower spectral type of the primary \citep{offner2023}.

\textsc{Hii} 1348A is almost an equal-mass binary \citep[$q = 0.7799 \pm 0.0098$;][]{torres2021}, however, we have elected to calculate the maximum observable photocenter shift possible for this binary to ensure that it does not significantly impact the uncertainty of our astrometric measurements where the stars are unresolved. The period of the binary is $94.805 \pm 0.012 \ \mathrm{d}$ \citep{torres2021}, which, together with the approximate masses of the binary stars, corresponds to a Keplerian semimajor axis of $0.43 \pm 0.01 \ \mathrm{au}$. The binary's orbital eccentricity is $0.554 \pm 0.002$ \citep{torres2021}. The maximum separation between the stars, $a(1+e) = 0.68 \pm 0.02 \ \mathrm{au}$, is less than half a pixel at the plate scale of our observations and for the distance to the \textsc{Hii} 1348 system. The maximum photocenter shift is much smaller than this maximum physical separation given the close spectral types---and therefore observed luminosities---of the two spatially unresolved binary components, along with the projection of the orbit plausibly reducing the observed size of the orbit.

The two binary host stars' $B-V$ colors are 1.05 and 1.35 for \textsc{Hii} 1348Aa and Ab, respectively. These colors correspond roughly to spectral types of K4V for \textsc{Hii} 1348Aa and K7V for \textsc{Hii} 1348Ab \citep[e.g.,][]{pecaut_mamajek_2013}. For this calculation, we have approximated \textsc{Hii} 1348Aa and Ab as 4620 K and 4050 K blackbodies, respectively, adopting the nominal effective temperatures corresponding to these approximate spectral types from \cite{pecaut_mamajek_2013}. Integrating over any one of the photometric bandpasses with reported measurements for \textsc{Hii} 1348B in Table \ref{tab:system}, the two stars would produce observed luminosities that differ by a factor $\approx 1.56$. Given the approximately equivalent inferred ratios of stellar fluxes in any bandpass, the maximum photocenter shift between observations during the host binary's apocenter in any one filter followed by observation at pericenter in any other filter, would imply identical maximum observed shifts.

Following a Keplerian analysis of the host binary stars' orbits, we have estimated the maximum photocenter shift across half the binary's orbital period ($P_\text{orb} / 2 = 47.403 \pm 0.012$ d), relative to the binary's center of mass. We find a maximum shift of $0.3 \pm 0.1 \ \mathrm{mas}$ in angular separation, or $0.02 \pm 0.01 \ \mathrm{px}$ imaged on the LMIRCam detector. This estimate ignores projection effects, assuming the maximum possible projected orbit size for zero inclination. The other sources of uncertainty present in the relative astrometry of \textsc{Hii} 1348B amount to ${\sim}5 \ \mathrm{mas}$, more than an order of magnitude larger than the estimated maximum photocenter shift. Therefore, we have chosen to neglect the astrometric uncertainty introduced by the host binary's photocenter shift across astrometric epochs.

$M\sin^3{i}$ measurements of the binary host stars are reported by \cite{torres2021} from long-term spectroscopic observations. With the same inferred masses from \cite{geißler2012} of $0.67 \pm 0.07 \ \mathrm{M_\odot}$ and $0.55 \pm 0.05 \ \mathrm{M_\odot}$, the two possible orbital inclinations of \textsc{Hii} 1348A are $i_\text{A} = 130^\circ \pm 2^\circ$ and $i_\text{A} = 50^\circ \pm 2^\circ$ (in the orbital basis of \texttt{orbitize!}) for clockwise (CW) or counter-clockwise (CCW) projected on-sky orbits of \textsc{Hii} 1348A, respectively.


\section{Discussion}

\subsection{Comparison to \cite{hurt_2024}}
\label{ssec:spec_fit_discussion}
 The effective temperature estimates for \textsc{Hii} 1348B of ${\sim}2900$ K that we find from \textsc{bt-settl} (CIFIST 2011/2015) atmospheric models, with an informed prior on $T_\text{eff}$, fall on the hot-end of our expectations for young ${\sim}$M8-type objects. Our prior assumption of effective temperature, from the spectral type of \textsc{Hii} 1348B compared with a polynomial relation published by \cite{sanghi_2023}, was $T_\text{eff} = 2500 \pm 300 \ \mathrm{K}$. The likelihood (i.e., the posterior with a uniform prior on effective temperature) selects for $T_\text{eff} \sim 3100$ K, which could be related to a possible deficiency in the treatment of dust opacity in \textsc{bt-settl} models when applied to M/L transition dwarfs for. This effect was extensively investigated by \cite{hurt_2024} for a large sample of young M/L transition dwarfs, including 26 isolated brown dwarfs in the Pleiades cluster. \cite{hurt_2024} find that \textsc{bt-settl} (CIFIST 2011/2015) models consistently overestimate the $J$- and $
H$-band fluxes for M/L transition dwarfs. Inspection of Figure \ref{fig:spectrum} indicates that the $J$-band flux of \textsc{Hii} 1348B is significantly overestimated by the maximum-a-posteriori models in either $\log(g)$-mode. However, the broadband $H$-band flux of \textsc{Hii} 1348B is well fit by these model atmospheres.

 Two types of systematic behavior for the atmospheric parameters inferred from fitting \textsc{bt-settl} (CIFIST 2011/2015) model spectra to late-M and L dwarfs are found by \cite{hurt_2024}. The first type corresponds to a group of objects clustered at inferred $T_\text{eff} \approx 1800$ K and $\log(g) \approx 5.5$ dex, implying very large masses of 150--1400 $\mathrm{M_J}$. The second type is a similar cluster of objects with inferred $T_\text{eff} \gtrsim 3000$ K and $\log(g) \lesssim 3.0$ dex, implying nonphysical, very low masses. This bimodal clustering behavior may be related to the bimodal likelihood and posterior distributions we find when applying the same model grid to photometry of \textsc{Hii} 1348B (see Figure \ref{fig:spectrum_post_surface}). One clear difference in the bimodal behavior found in this work from the clusters of \cite{hurt_2024} is our finding of a mode at $T_\text{eff} \sim 3000$ K with $\log(g) \approx 5.5$ dex---in addition to the $\log(g) \lesssim 3.0$ dex mode.

 \cite{hurt_2024} find seven ${\sim}\mathrm{M8}\text{--}\mathrm{L0}$ dwarfs clustered at $T_\text{eff} \gtrsim 3000$ K and very low $\log(g)$ in their analysis. Two of the seven are late-M ultracool dwarfs in the Pleiades: Roque 7 (M8.8) and UGCS J034334.48+255730.5 (M6.9). Such Pleiades late-M dwarfs would be expected to have very similar atmospheric properties to those of \textsc{Hii} 1348B---assuming that they may have likely formed via the same (or similar) mechanism(s). From their fitting of \textsc{bt-settl} (CIFIST 2011/2015) model spectra, \cite{hurt_2024} find $T_\text{eff} = 3027 \substack{+52 \\ -53}$ K with $\log(g) = 2.5 \substack{+0.25 \\ -0.24}$ dex for Roque 7, and $T_\text{eff} = 3084 \substack{+52 \\-53}$ K with $\log(g) = 2.51 \pm 0.25$ dex for UGCS J034334.48+255730.5. These values correspond well to the parameters of \textsc{Hii} 1348B nominally inferred from maximum-likelihood estimation of the low-$\log(g)$ mode seen in Figure \ref{fig:spectrum_post_surface}: $T_\text{eff} = 2970$ K with $\log(g) = 2.5$. The prior on $T_\text{eff}$ that we apply for MAP estimation serves to lower the temperature of this mode in the resultant posterior distribution (see Figure \ref{fig:spectrum_post_surface}). \cite{hurt_2024} investigate applying a broad prior on spectrally inferred mass, but find that it does not result in more accurate parameter estimation or particularly close fits of the models to their observations.  We encourage the $T_\text{eff}$ and $\log(g)$ inferred from our fitting to be interpreted with caution given the discussed behavior of the \textsc{bt-settl} (CIFIST 2011/2015) model grid as applied to the free-floating analogs of \textsc{Hii} 1348B in the Pleiades.

 The residual behavior for M/L transition objects fit by \cite{hurt_2024} with \textsc{bt-settl} spectra imply an under-accounting of dust opacity in such atmospheres, along with some opacity from FeH at $1.58 \ \mu\mathrm{m}$ possibly being excluded. It is probable that the inclusion of an extinction term similar to the ISM-like extinction investigated by \cite{hurt_2024}, or their suggestion of applying a model more appropriate to sub-micron-sized grains in brown dwarf atmospheres \citep[e.g.,][]{hiranaka_2016}, could improve the model fitting to \textsc{Hii} 1348B presented in this work. However, such an analysis would require the inclusion of additional free parameters to the fitting. With only six photometric measurements, such free parameters relating to extinction would add a significant penalty to the $\chi_\nu^2$ statistic. Therefore, we do not attempt to model the effects of dust extinction added to the \textsc{bt-settl} (CIFIST 2011/2015) atmospheric model grid.

 In the absence of a higher-resolution spectrum, we defer a more detailed analysis of the atmospheric properties of \textsc{Hii} 1348B to future studies. Additional infrared imaging and spectroscopy \citep[e.g., with LBTI/ALES;][]{skemer_2015, skemer_2018, hinz_2018, Stone2022} would aid further atmospheric characterization of this substellar companion. $M$-band imaging and JWST spectroscopy are just some of the possible avenues for further direct characterization of the atmosphere of \textsc{Hii} 1348B.

\subsection{System Dynamics and Stability}
With two sets of orbital information, one for the inner binary stars of \textsc{Hii} 1348A and another for the outer brown dwarf companion \textsc{Hii} 1348B, interesting insights into the dynamical history of this system may be drawn. The relatively high eccentricities of both the inner binary ($e_\mathrm{A} = 0.5543 \pm 0.0017$) and \textsc{Hii} 1348B ($e_\mathrm{B} = 0.78 \substack{+0.12 \\ -0.29}$) suggest a dynamical reprocessing of the three system components after their formation. The wide, likely highly eccentric orbit of \textsc{Hii} 1348B is consistent with a close three-body encounter with the two more massive host stars \citep[e.g.,][]{valtonen2006}. In particular, the much larger semimajor axis of the third component ($a_\mathrm{B} / a_\mathrm{A} = 330 \substack{+290 \\ -70}$) is consistent with a near-escape ejection of this brown dwarf companion. A close dynamical encounter of the system would explain the excited eccentricities of the inner binary and wide brown dwarf components and the clear hierarchy in orbital semimajor axes.

 While further orbital constraints would be necessary for accurate dynamical modeling of the system with an $N$-body integrator, some conclusions on the stability of the \textsc{Hii} 1348 system can be drawn. Long-term stability (i.e., a lack of collision or ejection of system components) for triple systems typically requires a hierarchical configuration---such as that of \textsc{Hii} 1348---consisting of an inner binary and an outer tertiary companion on nearly Keplerian orbits, with the tertiary's orbit being much wider than, and not making close approaches to, the inner binary \citep{naoz_2013b}. We infer the stability of this triple system with the $\epsilon$ criterion for hierarchical triple systems of \cite{naoz_2013b}, who demonstrate it to be numerically similar to the commonly used stability criterion of \cite{mardling_aarseth_2001} over a large range of masses,

 \begin{equation}
    \epsilon = \left(\frac{a_\text{A}}{a_\text{B}}\right) \frac{e_\text{B}}{1 - e_\text{B}^2}
\end{equation}\label{eq:triple_stab}
 where subscripts ``A'' and ``B'' refer to the orbits of the inner binary (A) and the outer tertiary companion (B). $\epsilon < 0.1$ is taken as a rule-of-thumb for system stability \citep{naoz_2016}. Note that this criterion does not consider the mutual inclination $i_\text{mut}$ of the inner and outer orbits (undetermined for the \textsc{Hii} 1348 system), where retrograde ($i_\text{mut} > 90^\circ$) configurations are expected to be more stable than prograde ($i_\text{mut} < 90^\circ$) configurations \citep[e.g.,][]{innanen_1979, innanen_1980, morais_giuppone_2012}. Using this criterion, we estimate a negligible $0.00058\%$ probability of system instability. To compute this estimate, we took the semimajor axis and eccentricity posterior distributions yielded by the \texttt{orbitize!}\ MCMC sampler together with an independent random sample of the same size ($10^7$) for the semimajor axis of the binary host stars. The sample of host-binary semimajor axes was drawn from a Gaussian distribution with mean 0.435 au and standard deviation 0.11 au, inferred from $a_\text{ab} \sin i$, $M_\text{a} \sin^3 i$, and $M_\text{b} \sin^3 i$ reported by \cite{torres2021} and the total binary mass of $M_\text{ab}$ estimated by \cite{geißler2012} (see Table \ref{tab:system}).

 From a sample of solar-type triple systems with orbital solutions, \cite{tokovinin_2004} provide an empirical stability relation based on the ratio of orbital periods $P_\text{B} / P_\text{A}$ of the system and the eccentricity of the outer component,

 \begin{equation}
    P_\text{B} / P_\text{A} > \frac{5}{(1 - e_\text{B})^3}
\end{equation}
 Applying this relation to the \textsc{Hii} 1348 system, we compute the distribution of inferred orbital periods of \textsc{Hii} 1348B from Kepler's Third Law: $P^2 = 4 \pi^2 a^3 / (GM_\star)$ \citep{kepler_1619, newton_1687}, and draw $10^7$ random samples for the period of the inner binary from a Gaussian distribution with mean 94.805 d and standard deviation 0.012 d \citep{torres2021}. This empirical criterion implies a less negligible, but yet very small, 1.1\% probability of system instability corresponding to the smallest and most highly eccentric orbits of \textsc{Hii} 1348B.

  The mutual inclination (i.e., the angle between the orbits' angular momenta), $i_\text{mut}$, between the inner binary and the orbit of \textsc{Hii} 1348B cannot be estimated due to the undetermined position angle of nodes ($\Omega_\text{A}$) for the host binary stars from spectroscopic observations alone. In addition to this unknown parameter, there are also degeneracies in the host-binary inclination $i_\text{A}$ and the companion's position angle of nodes $\Omega_\text{B}$, which allow for four equally likely values of $i_\text{mut}$ for any assumed value of $\Omega_\text{A}$. Therefore, all mutual inclinations between $0^\circ$ and $180^\circ$ are possible for the \textsc{Hii} 1348B system without additional constraining observations.

 The determination of a visual orbit for the inner host binary star with interferometric observations would allow for its inclination $i_\text{A}$ and position angle of nodes $\Omega_\text{A}$ to be uniquely identified via jointly fitting with the published RVs. Additionally, RV measurements of the wide orbit of \textsc{Hii} 1348B would break its joint $180^\circ$ degeneracy in $\Omega$ and $\omega$ (see Section \ref{sssec:orbit_fitting_results}). Therefore, if both a visual orbit of the host binary and a spectroscopic orbit of the wide companion were determined, the two orbits' orientations in 3-D space could be uniquely identified---allowing for a measurement of $i_\text{mut}$. Such a measurement would have implications on possible von Zeipel--Lidov-Kozai \citep[ZLK; ][]{vZ_1910, lidov1962, kozai1962} oscillation in the system, involving an exchange between mutual inclination and orbital eccentricity over secular timescales much larger than an orbital period.

 While we cannot measure $i_\text{mut}$ for \textsc{Hii} 1348 in this work, we may find some expectations from the results of studies analyzing samples of similar hierarchical triples. \cite{tokovinin_2017} found an equal number of prograde and retrograde configurations for triple systems with outer tertiaries at ${>}1000$ au, suggesting an isotropic distribution of $i_\text{mut}$ for such triple systems with very wide tertiaries. On the other hand, their study found that compact triple systems with outer components at ${ < }50$ au strongly favor prograde, relatively well-aligned configurations. We can confidently rule out such compact configurations for the \textsc{Hii} 1348 system, with none of the $10^7$ MCMC posterior orbits of \textsc{Hii} 1348B (see Section \ref{ssec:orbit_fitting}) having $a_\text{B} < 50$ au. However, the companion's orbit is also relatively unlikely to be as wide as those systems found by \cite{tokovinin_2017} to indicate isotropic $i_\text{mut}$, with only a 4.1\% probability of $a_\text{B} > 1000$ au. Consequently, orbital alignment would not be expected for this system, but it may still be more likely to possess a prograde (i.e., $i_\text{mut} < 90^\circ$) than a retrograde ($i_\text{mut} > 90^\circ$) configuration. The eccentric inner binary with $e_\text{A} = 0.5543 \pm 0.0017$ \citep{torres2021} may hint at misalignment, as \cite{tokovinin_2017} also find that, on average, better aligned hierarchical triple systems have lower inner binary eccentricities.

\newpage\subsection{Formation Pathways}
It appears most likely that \textsc{Hii} 1348B formed from the fragmentation of a disk or cloud core. It is possible that \textsc{Hii} 1348 formed as an initially bound triple system from the fragmentation of one protostellar disk \citep[e.g.,][]{kratter2010, bate2018, forgan2018} or from turbulent fragmentation of a molecular cloud \citep[e.g.,][]{bate2012, lee2019} followed by orbital migration. The observed orbital configuration does not provide strong evidence of the original fragmentation mechanism.  Within star clusters like the Pleiades, Jovian-to-super-Jovian planets and brown dwarfs can also be ``stolen'' from other systems or ``captured'' after existing as free-floating bodies in ${\lesssim}10 \ \text{Myr}$ \citep[e.g,][]{parker2022, daffern_powell_2022}. This allows for the possibility that \textsc{Hii} 1348B did not form together with its present host binary stars. In any case, a close dynamical encounter between the three known components of the \textsc{Hii 1348} system could produce the observed hierarchical configuration.

It is unlikely for \textsc{Hii} 1348B, with a mass of ${\sim}60\text{--}63 \ \mathrm{M_\mathrm{J}}$, to have formed through core accretion (CA). This process is not thought to form bodies larger than ${\sim}30 \ \mathrm{M_J}$ \citep[e.g.,][]{mordasini2009I, mordasini2009II, mordasini2012}. Further, \cite{schlaufman2018} reported evidence that the maximum mass of a planet formed via CA is ${\lesssim} 10 \ \mathrm{M_J}$. Additionally, the mass distribution of wide-orbit planets and brown dwarfs flattens out at masses greater than $10\text{–}20 \ \mathrm{M_J}$ \citep{wagner2019}. This is consistent with the scenario that CA is the primary mechanism of forming planets and brown dwarfs only at these masses or less, with \textsc{Hii 1348B} being $\gtrsim3\times$ more massive than this threshold. Gravitational instability (GI) in disks is most likely to form objects that orbit at ${\sim}20\text{–}60 \ \mathrm{au}$ \citep{forgan2013}, where the projected separation of \textsc{Hii} 1348B in 2019 was $> 150 \ \mathrm{au}$. Therefore, we infer that a close dynamical encounter in the system's history is required to explain the present orbit of \textsc{Hii} 1348B if this companion formed from the fragmentation of a protoplanetary disk.

\subsection{\textsc{Hii} 1348B as a Benchmark Substellar Object in the Pleiades}
With \textsc{Hii} 1348B being only one of three imaged substellar companions in the Pleiades, together with \textsc{Hii} 3441B and HD 23514B, it is placed at a unique age between imaged companions in younger moving groups and stellar associations, and those imaged around field-aged (gigayear-old) stars. At ${\sim}112 \ \mathrm{Myr}$, \textsc{Hii} 1348B is young enough to allow for constraints to be made on models of substellar cooling. Such ``hot-start'' models, with which almost all brown dwarfs are observed to be consistent, hot-start models \citep{spiegel2012} do not reach quasi-equilibrium until ages of several hundred megayears to ${\sim}1 \ \text{Gyr}$ \citep[e.g.,][]{baraffe2003}. Alternative ``cold-start'' models, e.g., \citep{marley2007}, assume lower initial internal entropies that should be more applicable to smaller Jovian planets formed via CA.

With the orbital modeling presented in this work, \textsc{Hii} 1348B is one of few known substellar companions with measured orbital motion beyond $100 \ \mathrm{au}$. Other very wide-orbit companions meeting this criterion include GQ Lupi B \citep{neuhauser2008, ginski2014, wu2017}, ROXs12 B \citep{bryan2016}, GSC 6214-210 B \citep{pearce2019} and HD 106906 b \citep{nguyen2021}. These very wide-orbit companions are all significantly younger than \textsc{Hii} 1348B: GQ Lupi is ${\sim}2\text{–}5$ Myr old, from spectropolarimetric observations \citep{donati2012}, The ROXs12 system has an estimated age of $7.6\substack{+4.1 \\ -2.5}$ Myr \citep{kraus2014}, GSC 6214-210 has an age of $16.9\substack{+1.9 \\ -2.9}$ Myr \citep{pearce2019}, and HD 106906 is a member of the 15 Myr--old Lower Centaurus--Crux subgroup of the Scorpius--Centaurus OB association \citep{pecaut2016}. Future astrometric modeling of the \textsc{Hii} 1348 system, especially with milliarcsecond precision, will further constrain the orbit of this brown dwarf and provide additional insights into the formation and dynamical history of the system.

\section{Conclusion}

Here we summarize our findings and conclusions:

\begin{enumerate}

\item We detected the known substellar companion to the spectroscopic binary star \textsc{Hii} 1348 in the $L^\prime$ filter for the first time.

\item We provide a new epoch of precise relative astrometric measurements, extending the existing temporal baseline of observations to 23 years.

\item We fit the orbit of \textsc{Hii} 1348B with an MCMC sampler, finding that it is likely to be on a very wide ($a = 140 \substack{+130 \\ -30} \ \mathrm{au}$), eccentric ($e = 0.78 \substack{+0.12 \\ -0.29}$) orbit.

\item  We fit \textsc{bt-settl} (CIFIST 2011/2015) model spectra \citep{allard2012, baraffe_2015} to all available apparent photometry of \textsc{Hii} 1348B, finding that the object is most consistent with $T_\text{eff} \sim 2900 \ \text{K}$, while its surface gravity remains unconstrained.

\item  From comparison of the estimated bolometric and absolute-$L^\prime$ luminosities of \textsc{Hii} 1348B to the evolutionary models of \cite{burrows_2001} and \cite{baraffe2003}, we infer \textsc{Hii} 1348B to have a mass of $60\text{--}63 \pm 2 \ \mathrm{M_J}$. The values that we obtain are consistent with earlier estimates, confirming the object's substellar nature.

\item Through synthetic source injections and comparison to the evolutionary tracks of \cite{baraffe2003}, we estimate mass detection limits of ${\sim}30 \ \mathrm{M_J}$ as close as 0\farcs{15} (${\approx}21.5$ au) to \textsc{Hii} 1348A, down to $10\text{–}11 \ \mathrm{M_J}$ in the background limit across projected separations of $0\farcs{25}\text{–}2''$ (${\approx}36\text{–}290 \ \mathrm{au}$).

\item We discuss the potential origin cases for \textsc{Hii} 1348B and the system's likely dynamical past. We infer that this companion likely formed through a large-scale gravitational instability, subsequently taking part in a dynamical interaction with the stars of the system's inner binary.

\end{enumerate}

In summary, we suggest that \textsc{Hii} 1348B is an important benchmark brown dwarf companion, adding valuable information for studying the population-level trends of wide-orbit substellar companions. Such objects may be the outliers of planetary and stellar formation models, and their observed orbital and spectroscopic properties allow for unique inferences on the dynamical interactions and formation processes which may explain their wide separations. The age of the \textsc{Hii} 1348 system at $112 \pm 5 \ \mathrm{Myr}$ \citep{dahm2015} also makes this object a unique laboratory for studying a young substellar companion that is yet older than those in the youngest (${\lesssim}20 \ \mathrm{Myr}$) stellar associations and moving groups which host the majority of such directly imaged companions to date.

\newpage\section{Acknowledgements}
\begin{acknowledgements}

We thank Amali Vaz, Emily Mailhot, and Jenny Power for assisting with the observations. We are also grateful for the anonymous referee's comments, which led to many improvements in this study. The results reported herein benefited from collaborations and/or information exchange within NASA’s Nexus for Exoplanet System Science (NExSS) research coordination network sponsored by NASA’s Science Mission Directorate. This material is based upon work supported by the National Aeronautics and Space Administration under Agreement No.\ 80NSSC21K0593 for the program ``Alien Earths.'' This research has made use of the SIMBAD database and VizieR catalogue access tool, CDS, Strasbourg, France. The LBT is an international collaboration among institutions in the United States, Italy, and Germany. LBT Corporation partners are: The University of Arizona on behalf of the Arizona university system; Istituto Nazionale di Astrofisica, Italy; LBT Beteiligungsgesellschaft, Germany, representing the Max-Planck Society, the Astrophysical Institute Potsdam, and Heidelberg University; The Ohio State University, and The Research Corporation, on behalf of The University of Notre Dame, University of Minnesota, and University of Virginia. Observations have benefited from the use of ALTA Center (\url{alta.arcetri.inaf.it}) forecasts performed with the Astro-Meso-Nh model. Initialisation data of the ALTA automatic forecast system come from the General Circulation Model (HRES) of the European Centre for Medium Range Weather Forecasts.  This research has made use of NASA's Astrophysics Data System. This publication makes use of data products from the Wide-field Infrared Survey Explorer, which is a joint project of the University of California, Los Angeles, and the Jet Propulsion Laboratory/California Institute of Technology, funded by the National Aeronautics and Space Administration. This publication makes use of data products from the Two Micron All Sky Survey, which is a joint project of the University of Massachusetts and the Infrared Processing and Analysis Center/California Institute of Technology, funded by the National Aeronautics and Space Administration and the National Science Foundation. This work has made use of data from the European Space Agency (ESA) mission Gaia (\url{https://www.cosmos.esa.int/gaia}), processed by the Gaia Data Processing and Analysis Consortium (DPAC, \url{https://www.cosmos.esa.int/web/gaia/dpac/consortium}). Funding for the DPAC has been provided by national institutions, in particular the institutions participating in the Gaia Multilateral Agreement.
\end{acknowledgements}
%

\facility{LBT (LMIRCam)}


\software{\texttt{astropy} \citep{astropy:2013, astropy:2018, astropy:2022}, \texttt{orbitize!}\ \citep{blunt2020}, \texttt{numpy} \citep{numpy}, \texttt{matplotlib} \citep{matplotlib}, \texttt{scipy} \citep{scipy}, \texttt{pandas} \citep{pandas}, \texttt{arviz} \citep{arviz}, and \texttt{SAOImageDS9} \citep{ds9}.}



\newpage\appendix
\section{Orbit-Fitting Corner Plot \& Discussion of Priors}\label{appendix:A}

\begin{figure*}[h]
\plotone{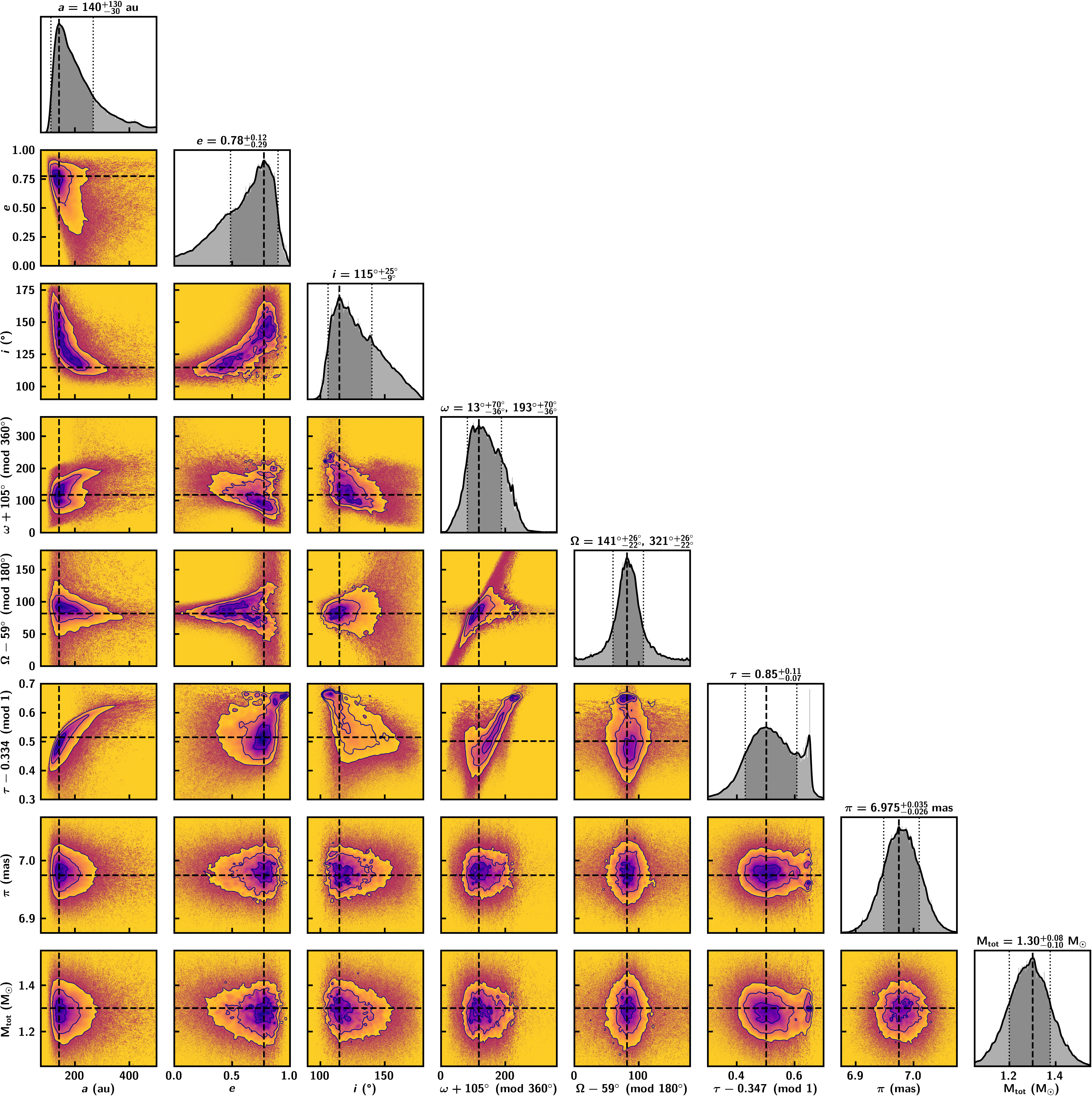}
\caption{Corner plot of the distribution of $10^7$ posterior orbit samples fitted with \texttt{orbitize!}/MCMC (see Section \ref{ssec:orbit_fitting}). Diagonal subplots show 200-bin histograms of 1-D marginalized distributions with KDEs over-plotted in solid black. Dashed and dotted lines mark MAP estimates and bounds of 68\% HPD regions. Solid contours in off-diagonal subplots mark 1.5-, 1-, and 0.5-$\sigma$ levels containing ${\sim}67.5$\%, ${\sim}39.3$\%, and ${\sim}11.8$\% posterior probability. Contours are estimated from 2-D KDEs, with samples scattered outside of the 1.5-$\sigma$ level. The distributions of $\omega$, $\Omega$, and $\tau$ have been shifted by constant offsets for readability. The marginalized distributions of $\omega$ and $\Omega$ show only one of two such degenerate modes present in the absence of RV constraints. MAP estimates and 68\% HPD regions corresponding to both degenerate modes are annotated (separated by 180\textdegree) above the diagonal for these parameters.
\label{fig:corner_plot}}
\end{figure*}

 Recent analyses of orbit fitting in Bayesian frameworks have shown that the common adoption of uniform priors on some orbital parameters (e.g., $\tau$ and $e$) used in this analysis may introduce unintended biases to, and under-estimate the credible regions of, the posterior when limited orbital phase is observed \citep[e.g.,][]{ferrer_chavez_2021, konopacky_2016, oneil_2019, doo2023}. These effects can be understood as resulting from the uniform priors dominating over the likelihood where limited astrometric data is available to constrain the posterior \citep{oneil_2019}.

Given the short observed orbital coverage for \textsc{Hii} 1348B, we therefore advise some caution in interpreting the MAP estimates and HPD regions presented in this work. ``Observable-based'' priors---introduced by \cite{oneil_2019} and applied by \cite{doo2023} to an analysis of the population-level eccentricity distribution of directly imaged exoplanets---are an alternative set of orbital priors which are designed to reduce their contribution to the posterior, thereby increasing the contribution of the observations. The \texttt{orbitize!}\ Python package, at the time of writing, supports the implementation of such observable-based priors, though without the functionality to fit for total system mass or parallax as free parameters. Continued orbital modeling of \textsc{Hii} 1348B will provide a more accurate and precise orbit determination for this substellar companion independent of prior assumptions. This could reasonably include additional relative astrometry from high-contrast imaging obtained at future epochs, and/or the inclusion of RV observations to simultaneously fit for the inner and outer orbits of \textsc{Hii} 1348.

\section{Fitting to \textsc{bt-settl} Models with Non-Solar Abundances}
\label{appendix:B}

 In our first attempt to fit spectral models to the available photometry of \textsc{Hii} 1348B, we explored a larger grid of \textsc{bt-settl} models with variation in metallicity and alpha enhancement. We found that the metallicity and alpha enhancement were ill-constrained by broadband photometry of \textsc{Hii} 1348B alone. Inspection of the relatively broad posterior showed that metallicities around solar values were generally preferred. With this in mind, we elected to restrict the grid of atmospheric models to the \textsc{bt-settl} (CIFIST 2011/2015) grid, which fixes abundances to solar values found by \cite{caffau_2011} and offers some improvements, detailed in \cite{baraffe_2015}, which are not included in other \textsc{bt-settl} grids. Additionally, we were motivated to restrict our search to solar-abundance atmospheres by measurements of the mean metallicity of the Pleiades cluster being very close to solar, e.g., \cite{grilo_2024} report an average cluster metallicity of $[\mathrm{Fe}/\mathrm{H}] = +0.03 \pm 0.04$.

\bibliography{HII_1348_B}{}
\bibliographystyle{aasjournal}

\end{document}